\documentclass[10pt]{IEEEtran}
\usepackage{float}
\usepackage{amsmath}
\usepackage{amsthm}
\usepackage{amssymb}	
\usepackage{graphicx}
\usepackage{subfigure}
\usepackage{enumitem}
\usepackage{algorithm}
\newcommand*{\J}{\jmath}%
\usepackage{amsmath,amssymb,mathtools,bm,etoolbox}
\newtheorem{my_theorem}{Theorem}

\newtheorem{my_lemma}{Lemma}

\usepackage{color}
\usepackage{mathtools}
\usepackage{suffix}
\usepackage{breqn}
\usepackage{hhline}
\usepackage{cite}

\DeclarePairedDelimiterXPP\Aver[1]{\mathbb{E}}{[}{]}{}{
	
	#1
}
\DeclarePairedDelimiterX\MeijerM[3]{\lparen}{\rparen}%
{\,#3\delimsize\vert\begin{matrix}#1 \\ #2\end{matrix}}
\newcommand\MeijerG[8][]{%
	G^{\,#2,#3}_{#4,#5}\MeijerM[#1]{#6}{#7}{#8}}
\newcommand{\diff}{\mathop{}\!d}

\WithSuffix\newcommand\MeijerG*[7]{%
	G^{\,#1,#2}_{#3,#4}\MeijerM*{#5}{#6}{#7}}

\title{Reconfigurable Intelligent Surface for Vehicular Communications: Exact Performance Analysis with Phase Noise  and Mobility} 
\author{Vinay Kumar Chapala,~\IEEEmembership{Graduate Student Member,~IEEE} and  S.~M.~Zafaruddin,~\IEEEmembership{Senior Member,~IEEE} 
		\thanks{A preliminary version of the paper without phase noise and mobility with symmetric dGG fading channel was presented at the 2022 IEEE 95th Vehicular Technology Conference (VTC2022-Spring), Helsinki, Finland, 19-22 June 2022 \cite{chapala2022vtcdgg}.}
			\thanks{ This work was supported in part by the Science and Engineering Research Board (SERB), Government of India, under MATRICS Grant MTR/2021/000890 and Start-up Research Grant SRG/2019/002345.}	
	\thanks{Vinay Kumar Chapala (p20200110@pilani.bits-pilani.ac.in) and S.~M.~Zafaruddin (syed.zafaruddin@pilani.bits-pilani.ac.in)  are  with  the Department of Electrical and Electronics Engineering, Birla Institute of Technology and Science, Pilani, Pilani Campus,  Pilani-333031, Rajasthan, India.}

}

\begin{document}
	\maketitle
	\begin{abstract}
Recent research provides an approximation on the performance for reconfigurable intelligent surface (RIS) assisted systems over generalized fading channels with phase noise resulting from imperfect phase compensation at the RIS. This paper presents an exact analysis of RIS-assisted vehicular communication by coherently combining received signals reflected by RIS elements and direct transmissions from the source terminal, considering generalized fading channels and phase noise. We adopt the generalized uniform distribution to model the phase noise resulting from a finite set of phase shifts at the RIS. We use generalized-$K$ shadowed fading distribution for the direct link and asymmetrical channels for the RIS-assisted transmission with $\kappa$-$\mu$ distribution for the first link and double generalized Gamma (dGG) distribution for the second link combined with a statistical random way point (RWP) model for the destination under movement. We employ a novel approach to represent the probability density function (PDF) and cumulative distribution function (CDF) of the product of four channel coefficients in terms of a single univariate Fox's H-function, and develop an exact statistical analysis of the  end-to-end signal-to-noise ratio (SNR) for the considered system using multi-variate Fox's H-function. We also use the inequality between the arithmetic and geometric mean to simplify the statistical results of the considered system in terms of the univariate Fox's H-function. We present exact, upper bound,  and asymptotic expressions of the outage probability and average bit-error-rate (BER) performance using the derived density and distribution functions.   We validate the analytical expressions through numerical and simulation results and demonstrate the effect of phase noise and mobility on the  performance scaling with RIS elements and direct transmissions considering various practically relevant scenarios.
		
	\end{abstract}

	
	\section{Introduction}

\begin{table*}[tp]
	\caption{Related Research With Phase Noise}
	\label{table:ris_phase_comp}
	\centering
	\begin{tabular}{ c c c c c c }
		\hline
		\hline
		Ref   & Direct Link &   Mobility & RIS Channels &  Fading Types & Analysis\\
		\hline
		[23] &  No & No & Symmetric & (Rayleigh, Rayleigh) & CLT \\  \hline
		[29] & No & No &  Symmetric &  (General, General) & Approximation \\ \hline
		[24] & Yes & No & Symmetric &  (Rayleigh, Rayleigh) & CLT \\ \hline
		[25] & Yes & No & Symmetric &  (Rayleigh, Rayleigh) & CLT \\ \hline
		[27] &  No &No & Symmetric & (Rician, Rician) & Approximation \\ \hline
		[22] & No & No &Symmetric& (Fox-H, Fox-H) & CLT, lower and upper bounds \\ \hline
		[28] & No & No & Symmetric &  (Nakagami-m, Nakagami-m) & Exact for integer $m$ \\ \hline
		This Paper & Yes & Yes  & Asymmetric &  ($\kappa$-$\mu$, dGG) & Exact and Bounds\\
		\hline
		\hline
		
	\end{tabular}
\end{table*}

The upcoming 6G wireless system envisions to cater exceedingly higher requirements of network throughput,  lower latency of transmission, and stable  connectivity  for autonomous vehicular communications \cite{Yishi_2021}.  However, dynamic channel conditions between an access point to a  vehicle in motion may be a bottleneck for the desired quality of service. The use of 	reconfigurable intelligent surface  (RIS) can be a promising technology to create a strong line-of-sight (LOS) transmissions by artificially controlling the characteristics of propagating signals for vehicular communications \cite{Yishi_2021, Qingqing2021,Basar2019_access}. An RIS is a planar metasurface made up of a large number of inexpensive passive reflective elements that can be electronically programmed to customize the phase of the incoming electromagnetic wave for enhanced signal quality and transmission coverage. 

At the outset,  related research  assumed ideal phase compensation at the RIS module to analyze the performance of  RIS empowered  wireless networks such as radio-frequency (RF)  \cite{Kudathanthirige2020, Qin2020, Ferreira2020,   Selimis2021, Khoshafa,trigui2020_fox, du2021}, free-space optics (FSO)   \cite{Jamali2021, chapala2021unified},  terahertz (THz) \cite{du2020_thz, chapala2021THz}, and  vehicular  \cite{Wang_2020_outage,Kehinde_2020,dampahalage2020intelligent,Long_2021, Ozcan2021,Abubakar_2020}. Under the assumption of the ideal phase, statistical characterization of the RIS-assisted system requires the derivation  for the sum of the product of two fading coefficients corresponding to the two links between the source to the RIS and RIS to the destination.   Initially, the research works  applied  the central limit theorem (CLT) to develop approximate analysis for various fading models \cite{Kudathanthirige2020,Qin2020,Ferreira2020,  Selimis2021,Wang_2020_outage}.   Recently, the authors in \cite{trigui2020_fox, du2020_thz,chapala2021THz,chapala2021unified} used the multivariate Fox's H-function representation to develop an exact analysis of various RIS-assisted wireless networks.  The authors in \cite{trigui2020_fox} presented an exact performance analysis for RIS-assisted wireless transmissions over generalized Fox’s H fading channels. The authors in \cite{du2020_thz, chapala2021THz}  analyzed  RIS-aided THz communications over different fading channel models combined with antenna misalignment and hardware impairments. In \cite{chapala2021unified}, an  unified  performance analysis for a FSO system was presented considering different atmospheric turbulence models and pointing errors.  In the conference version of the paper, we presented an exact analysis for RIS-aided transmissions with direct link over double generalized (dGG) fading  that accurately models the vehicular communications  \cite{chapala2022vtcdgg}.

It should be emphasized that the above and related research assumed ideal phase compensation at the RIS to facilitate performance analysis. However,  phase noise resulting from imperfect phase compensation at the RIS is inevitable in a practical setup. The impact of phase noise becomes of a paramount importance, especially in vehicular communications considering the dynamics of channel between the RIS and the destination under movement.  Generally, the phase noise is  statistically distributed as  Gaussian or generalized uniform \cite{Trigui_phasenoise}. Under the statistical phase noise,   performance of RIS-assisted wireless system requires  the analysis for the sum of the product of three fading coefficients. As is for perfect phase compensations, existing works  approximate the performance of RIS system for various fading channels  in the presence of phase noise \cite{Kudathanthirige2020, Qian_RIS_phase,LiDong_RIS_phase, wang_RIS_phase, Peng_RIS_phase,waqar2021performance,Tegos_RIS_phase}. In \cite{Kudathanthirige2020}, \cite{Qian_RIS_phase}, \cite{LiDong_RIS_phase}, \cite{wang_RIS_phase}, authors  applied the CLT to analyze the performance of RIS system  over Rayleigh fading channel with phase errors. The authors in \cite{Badiu_RIS_phase} approximated an  arbitrary fading model with Nakagami-$m$ distributed to develop performance analysis of the RIS-assisted transmission with phase noise. The reference \cite{Peng_RIS_phase} studied the effect of quantization level of the phase noise on  the diversity order of the system. The authors in \cite{waqar2021performance} derived approximate performance bounds for Rician fading model with phase errors.

It is desirable to provide an exact  analysis for RIS-assisted system in the presence of phase noise for a better performance assessment.  In \cite{Tegos_RIS_phase}, the authors derived exact statistical analysis for RIS-assisted system over Nakagami-$m$ fading channel  with phase noise. However, the analysis is valid for integer values of fading parameter $m$ and may approximate the performance of the system when the parameter $m$ becomes real. Similar to the case for perfect phase compensation, it is natural to apply the multivariate Fox's H-function to represent the analysis with the phase noise in an exact form. However, an application of the theory of random variables for the product of three channel coefficients with generalized fading models results into  bivariate Fox's H-function. However,  statistical representation of  the sum of bivariate Fox's H-functions is intractable  mathematically. Recently, the authors in \cite{Trigui_phasenoise} provided an exact analytical expression using multivariate Fox's H-function for RIS-assisted system with generalized fading models without phase noise but  provided approximate analysis in the presence of phase noise. For vehicular communications, the phase noise at the RIS module should not be neglected considering the dynamics of the channel due to the mobility effect.  Moreover, the two links from source to RIS and RIS to destination may not incur symmetrical fading. Furthermore, direct link with the combined effect of fading and shadowing should be included in the analysis.  To the best of the author's knowledge, an exact analysis of RIS-assisted wireless transmission with phase noise and user mobility over asymmetrical and generalized fading channels considering the shadowed direct link  has not been studied yet. A summary of related research on the topic has been provided in Table 	\eqref{table:ris_phase_comp}.

In this paper, we analyze the performance of RIS-assisted vehicular communication system by coherently combining received signals reflected by RIS elements and direct transmissions from the source terminal, considering generalized fading channels and phase noise. The major contributions of the paper are listed as follows:
\begin{itemize}
\item  We adopt the generalized uniform distribution to model statistically the phase noise resulting from a finite set of phase shifts at the RIS. 
\item We use generalized-$K$ shadowed fading distribution for the direct link and asymmetrical channels for the RIS-assisted transmission with $\kappa$-$\mu$ distribution for the first link and double generalized Gamma (dGG) distribution for the second link combined with a statistical random way point (RWP) model for the destination under movement.
\item  We employ a novel approach to represent the probability density function (PDF) and cumulative distribution function (CDF) of the product of four channel coefficients in terms of a single univariate Fox-H function. We use the derived PDF to develop an exact statistical analysis of the  end-to-end signal-to-noise ratio (SNR) for the RIS-assisted system using multi-variate Fox-H function.  Further,  we use the inequality between the arithmetic and geometric mean to simplify the statistical results of the considered system in terms of the univariate Fox-H function.
\item  We present exact, upper bound,  and asymptotic expressions of the outage probability and average bit-error-rate (ABER) performance using the derived density and distribution functions  for  better Engineering insights on the system performance at a high SNR. 
\item We validate the analytical expressions through numerical and simulation results and demonstrate the effect of phase noise and mobility on the  performance scaling with RIS elements and direct transmissions considering various practically relevant scenarios.
\end{itemize}

	\subsection*{Description of Main Notations:}
Main notations used in this paper are as follows: $\J$ denotes the imaginary number, $\Aver{.}$ denotes the expectation operator, $\exp(.)$ denotes the exponential function, while $\Gamma(a)=\int_{0}^{\infty}u^{a-1} e^{-u} \diff u$ denotes the Gamma function. We denote the Meijer's G-function by $G_{p,q}^{m,n}  \left(x \middle\vert \begin{array}{c} \{a_{w}\}_{w=1}^{p}\\ \{b_{w}\}_{w=1}^{q} \end{array} \right)$, the single-variate Fox's H-function by $H_{p,q}^{m,n}  \left(x \middle\vert \begin{array}{c} \{(a_{w},A_{w})\}_{w=1}^{p}\\ \{(b_{w},B_{w})\}_{w=1}^{q} \end{array}\right)$,  $N$-variate Fox's H-function as defined in \cite[A.1]{M-Foxh} and a shorthand notation  $\{a_{i}\}_{1}^{N} = \{a_{1},\cdots,a_{N}\}$ to represent their coefficients.  
	\subsection*{Organization of the Paper:}
This paper is organized as follows: In Section II, we describe the system  with channel fading models. In Section III, we develop statistical results for RIS-assisted transmission with phase noise and mobility. In Section IV, we analyze the performance of considered system using outage probability and average BER.  In Section V, numerical and simulation results are presented. Section VI concludes the paper.

\section{System Model}	
\begin{figure}
	\centering
	\vspace{-26mm}
	\includegraphics[scale=0.35]{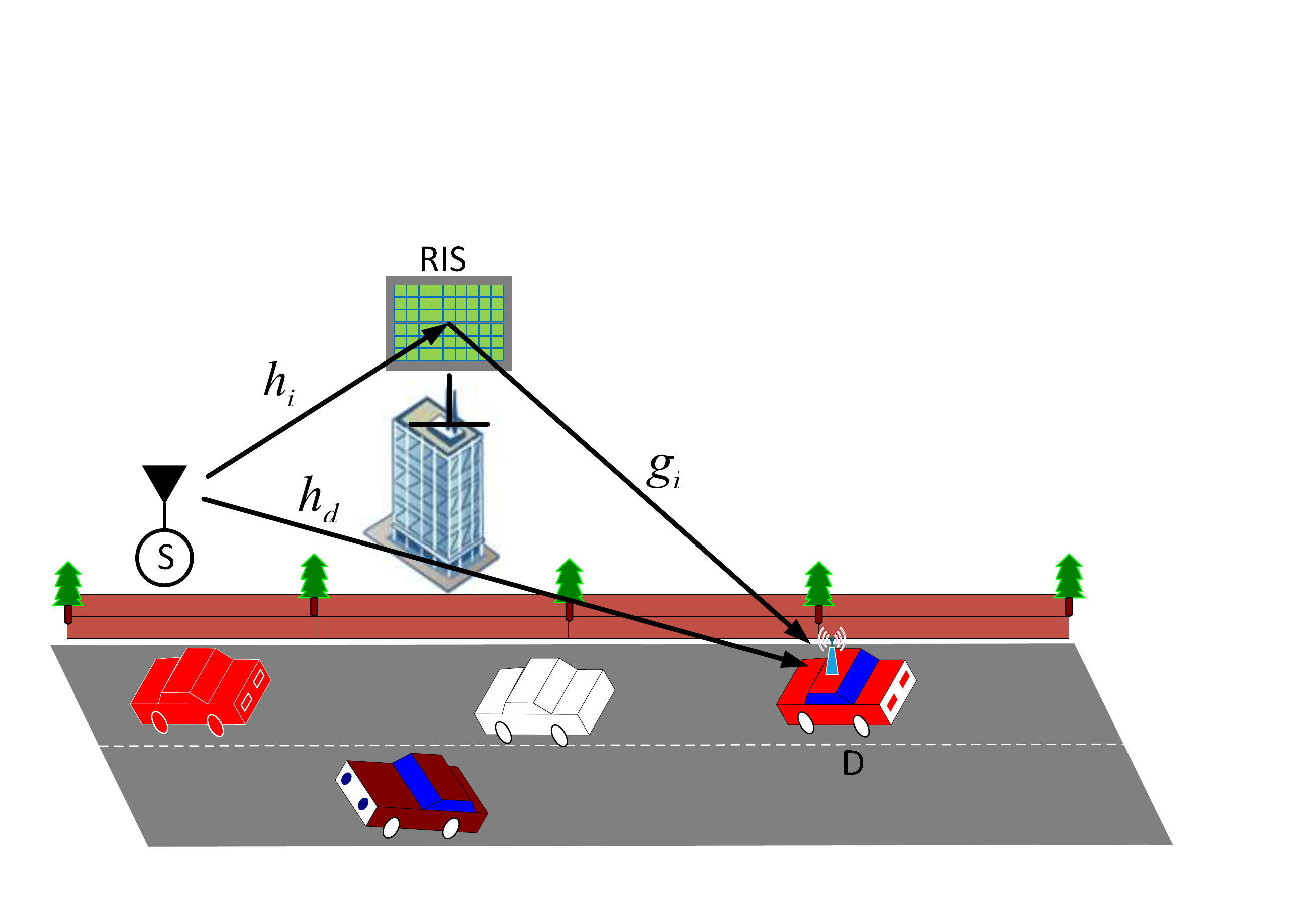} 
	\caption{System Model.}\vspace{-4mm}
	\label{fig:system_model}
\end{figure}

	We consider a  transmission model, where a single antenna source communicates with a destination equipped with a single antenna through an RIS with $N$ reflecting elements as shown in Fig.~\ref{fig:system_model}. We consider that the  elements of RIS are  spaced half of the wavelength and assume independent channels at the RIS \cite{Basar2019_access, du2021, Dovelos2021}.  We denote $h_{i}^{(f)}=\lvert h_{i}^{(f)}\rvert e^{j\theta_{h_i}}$  and  $g_{i}^{(f)}=\lvert g_{i}^{(f)}\rvert e^{j\theta_{g_i}}$  are channel fading coefficients between the source to the $i$-th RIS element and between the $i$-th RIS element to the destination, respectively. Assuming flat fading channel and imperfect phase compensation, the residual phase is non-zero, $\phi_{i}=-\theta_{h_{i}}-\theta_{g_{i}}+\theta_{i}$, where $\theta_{i}$ is random phase noise, $\theta_{h_i}$ is the phase of the first link, and $\theta_{g_i}$ is the phase of the second link. Thus, the received signal becomes
\begin{eqnarray}\label{model_phase_noise}
	y = \sqrt[]{P_{t}} x \left( \sum_{i=1}^{N} H_{l,i} h_{i}^{(f)}  g_{l,i}^{\rm rwp}  g_{i}^{(f)}  e^{\J \phi_{i}}+ \omega h_{l,d}^{\rm rwp}  h_{d}^{(f)}  \right) + v
\end{eqnarray}
where $P_{t}$ is the transmit power, $x$ is the unit power information bearing signal, $\lvert h_{i}^{(f)}\rvert$  and  $\lvert g_{i}^{(f)}\rvert$  are channel fading coefficients between the source to the $i$-th RIS element and between the $i$-th RIS element to the destination, respectively, $\lvert h_d^{(f)}\rvert$ is the flat fading coefficient between source and destination, and $v$ is the additive Gaussian noise with zero mean and variance $\sigma_v^2$. The terms $H_{l,i}$, $g_{l,i}$ and $h_{l,d}$ denote the path loss coefficients from source to RIS, RIS to destination and source to destination respectively. We denote $g_{i}=g_{l,i}\lvert g_{i}^{(f)}\rvert$ and for direct link, $h_{d}=h_{l,d}\lvert h_{d}^{(f)}\rvert$. Here, $\omega=1$ depicts the scenario with direct link and $\omega=0$ corresponds to the case without the direct link. Let $d_{1}$ and $d_{2}$ be the distances from source to RIS and RIS to destination. The distance from source to destination is considered as $d=\sqrt{d_{1}^{2}+d_{2}^{2}}$. The path loss component from source to RIS is considered as $H_{l,i}=d_{1}^{-\frac{a}{2}} (\frac{c}{4\pi f_{c}})^{}$, where $a$ is the path loss exponent, $c$ is the speed of the light and $f_{c}$ is the carrier frequency.

\subsection{Fading Models}
We assume that RIS is placed close to the source and destination is relatively far from the RIS and hence both the links may undergo asymmetrical fading. Most of the existing studies model the RF link as Rayleigh, Nakagami-$m$, or Rician fading. These channel models may not be an actual representation of fading characteristics of the RF link. The use of line-of-sight (LOS) based fading models are more suitable for RIS transmission. The $\kappa$-$\mu$ fading model is a generalized LOS model    fits well with experimental data over a wide range of scenarios \cite{Yacoub_kmu_etamu}. The $\kappa$-$\mu$  contains Nakagami-m ($\kappa \to 0$, $\mu=m$) model and the well studied  Rician ($\mu=1$) fading channel for RIS-assisted transmission as a particular case.  

We consider the channel from source to RIS $\lvert h_{i}^{(f)}\rvert$ distributed as $\kappa$-$\mu$ with the PDF \cite{Yacoub_kmu_etamu}
\begin{eqnarray}
	\label{eq:hi_pdf}
	&f_{\lvert h_{i}^{(f)}\rvert}(x) = \frac{2\mu_{i} (1+\kappa_{i})^{\frac{1+\mu_{i}}{2}} x^{\mu_{i}}}{\kappa_{i}^{\frac{\mu_{i}-1}{2}} {\rm exp}(\mu_{i} \kappa_{i})} {\rm exp}\Big[-\mu_{i} (1+\kappa_{i})x^2\Big] \nonumber \\ &I_{\mu_{i}-1}\Big[2\mu_{i} \sqrt{\kappa_{i}(1+\kappa_{i})} x\Big]
\end{eqnarray} 
where $\kappa_{i}$ is the ratio of power of the dominant components and the total power of scattered waves and $I_{v}(x)$ is the modified Bessel function of the first kind of order $v$ and can be represented by its infinite series as $I_{v}(x) = \sum_{k=0}^{\infty} \frac{1}{k! (v+k+1)} (\frac{x}{2})^{v+2k}$. Using the series expansion and after some algebraic manipulations, we get
\begin{eqnarray}
	\label{eq:hi_pdf_simp}
	&f_{\lvert h_{i}^{(f)}\rvert}(x) = \sum_{k_{i}=0}^{\infty}\psi_{i,1} x^{\mu_{i}+k_{i}-1} e^{-\zeta_{i,1} x}
\end{eqnarray}
where $\psi_{i,1}=\frac{1}{k_{i}! (\mu_{i}+k_{i})} \frac{\mu_{i}^{\mu_{i}+2k_{i}} \kappa_{i}^{k_{i}} (1+\kappa_{i})^{\mu_{i}+k_{i}} }{{\rm exp}(\mu_{i} \kappa_{i})}$ and $\zeta_{i,1}=\mu_{i}(1+\kappa_{i})$.
We model the channel between the RIS to the destination $\lvert g_{i}^{(f)}\rvert$ distributed as double Generalized-Gamma (dGG) with the PDF \cite{chapala2022vtcmultiris}:
\begin{eqnarray}\label{eq:dgg_pdf_new}
	&f_{\lvert g_i^{(f)}\rvert}(x) = \frac{1}{\Gamma(\beta_{i,1})\Gamma(\beta_{i,2})} x^{-1} \nonumber \\ &H_{0,2}^{2,0} \Bigg[\begin{array}{c} \phi_{g,i}^{\frac{1}{\alpha_{i,2}}} x^{} \end{array} \big\vert \begin{array}{c} - \\ (\beta_{i,2},\frac{1}{\alpha_{i,2}}),(\beta_{i,1},\frac{1}{\alpha_{i,1}})\end{array}\Bigg]
\end{eqnarray}
where $\phi_{g,i}=\frac{\beta_{i,2}}{\Omega_{i,2}} \big(\frac{\beta_{i,1}}{\Omega_{i,1}}\big)^{\frac{\alpha_{i,2}}{\alpha_{i,1}}}$, $\alpha_{i,1}$, $\beta_{i,1}$, $\alpha_{i,2}$, $\beta_{i,2}$ are Gamma distribution shaping parameters and $\Omega_{i,j}=\big(\frac{\mathbb{E}[\chi_{i,j}^{2}]\Gamma(\beta_{i,j})}{\Gamma(\beta_{i,j}+2/\alpha_{i,j})}\big)^{\alpha_{i,j}/2}$, $j=1,2$ is the $\alpha_{i,j}$-root mean value.
 The dGG model is suitable for modeling non-homogeneous double-scattering radio propagation fading conditions for vehicular communications over RF frequencies \cite{Petros2018}.  The dGG is a generalized model which  contains most of the existing statistical models such as double-Nakagami, double-Rayleigh, double Weibull, and Gamma-Gamma as special cases. 

We use the generalized $\mathcal{K}$ distribution to model the direct channel $\lvert h_{d}^{(f)}\rvert$ from $S$ to $D$ under  the combined effect of short term fading  and shadowing  \cite{Shankar_GenK}
\begin{eqnarray}\label{eq:hd_pdf}
	f_{\lvert h_d^{(f)}\rvert}(x) = \frac{2b}{\Gamma(m)\Gamma(M)} \bigg(\frac{bx}{2}\bigg)^{M+m-1} K_{M-m}(bx)
\end{eqnarray}
where $b=2\sqrt{\frac{m}{m_{0}}}$ and $K_{M-m}(.)$ is the modified bessel function of second kind of the order $(M-m)$. Here, $M = \frac{1}{{\rm exp}(\sigma_{n}^2)-1}$, $\sigma_{n}$$>$$0$ denotes the severity of shadowing with factor $\sigma_{\rm dB} =8.686\sigma_{n} $, $m$$>$$0$ is frequency dependent parameter characterizing the small-scale fading, and $m_{0}$ is a measure of mean power.

We express $K_{M-m}(bx)$ in terms of Meijer G-function as $K_{M-m}(x) = \frac{1}{2} G_{0,2}^{2,0} \bigg(\begin{array}{c} \frac{b^{2} x^{2}}{4} \end{array} \big\vert \begin{array}{c} - \\ \frac{M-m}{2},\frac{m-M}{2}\end{array}\bigg)$ to get
\begin{eqnarray}\label{eq:hd_pdf_1}
	&f_{\lvert h_d^{(f)}\rvert}(x) = \frac{b}{\Gamma(m)\Gamma(M)} \bigg(\frac{bx}{2}\bigg)^{M+m-1} \nonumber \\ &G_{0,2}^{2,0} \bigg[\begin{array}{c} \frac{b^{2} x^{2}}{4} \end{array} \big\vert \begin{array}{c} - \\ \frac{M-m}{2},\frac{m-M}{2}\end{array}\bigg]
\end{eqnarray}

\subsection{Statistical Models for Phase Noise and Mobility}
 In the literature, there are three different models, Gaussian, generalized uniform and uniform RVs for phase noise \cite{Trigui_phasenoise}. In the practical setup, due to hardware limitations, only a finite set of phase shifts can be realized that lead to quantization errors. To this extent, in our analysis, we consider generalized uniform distribution i.e., $\theta_{i}\sim\mathcal_{U}(-q\pi,q\pi)$ that models the quantization noise for $q=2^{-L}$, where $L\geq1$ denotes the number of quantization bits. It should be noted that assumption of perfect phase compensation may not be appropriate for vehicular communications.

We use the RWP model to consider the effect of mobility in the RIS to destination link \cite{Govindan_mobility}. The average received power in general depends on the propagation distance as $P_{t}r^{-a}$, where $P_{t}$ is the transmit power and $a$ $(2 \le a\le 5)$ is the path-loss exponent that depends on the propagation environment. For the static user, the path loss component $r^{-a}$ is constant but for a moving user, the distance $r$ is a random variable. In the RWP mobility model, it is usually assumed that the receiving users are located at randomly selected coordinate points in the vicinity of the transmitter, which depends on the network topology. For a $1$-D topology, we consider a line with the transmitter being located at the origin. The 2-D topology is assumed to be a circle, whereas a 3-D topology is a spherical network. In both the $2$-D and $3$-D network topologies, it is assumed that the transmitter is located at the origin. The RWP mobility models are polynomials in the transmitter–receiver distance $r$, and the PDF of the distance $r$ is given by the general form \cite{Govindan_mobility}

\begin{eqnarray}\label{eq:mobility_power_pdf}
	f_{r}(x) = \sum_{j=1}^{n} B_{j} \frac{r^{\beta_{j}}}{d_{}^{\beta_{j}+1}}
\end{eqnarray}
where parameters $n$, $B_{j}$, and $\beta_{j}$ depend on the number of dimensions considered in the topology  summarized in \cite{Govindan_mobility}. For example, \cite[Table  I]{Govindan_mobility} gives $n = 2$,$B_{j} = [6,-6]$, and $\beta_{j} = [1,2]$ for 1-D topology; $n = 3$,$B_{j} = (\frac{1}{73}) \times [324,-420,96]$, and $\beta_{j} = [1,3,55]$ for 2-D topology; and $n = 3$,$B_{j} = (\frac{1}{72}) \times [735,-1190,455]$, and $\beta_{j} = [2,4,6]$ for 3-D topology. The authors in \cite{Kehinde_2020,Sikri_RIS_mob} also employed RWP model for analyzing the performance of RIS system for the user under mobility.

\section{Statistical Derivations}
In this section, we develop probability density and distribution functions for the effective channel for  RIS-assisted transmission with phase noise and mobility. Without the loss of generality, we assume $H_{l,i}= H_{l}, \forall i$, and rewrite \eqref{model_phase_noise} as: 
\begin{eqnarray}\label{model_phase_noise_z}
y = \sqrt[]{P_{t}}   x\left( H_{l}\sum_{i=1}^{N} Z_i+ \omega Z_d  \right) + v
\end{eqnarray}
where $Z_i=\lvert h_{i}^{(f)}\rvert  g_{l,i}^{\rm rwp} \lvert g_{i}^{(f)}\rvert  e^{\J \theta_{i}}$ is the product of four random variables and $Z_d=h_{l,d}^{\rm rwp} \lvert h_{d}^{(f)}\rvert$ is the product of two random variables.
\subsection{Without Direct Link ($\omega=0$)}
We use the multivariate Fox's H-function to develop exact expressions of the PDF and CDF for $Z_{\rm RIS}= \sum_{i=1}^{N} Z_i$. We also present bounds on analytical expressions for RIS-assisted transmission using univariate Fox's H-function for better insights into the system performance. 

Since $Z_i$ is the product of four random variables with complicated distribution functions, a straight forward application of  the probability theory may lead the PDF of $Z_i$ in terms bivariate/trivariate  Fox's H-function, precluding the representation of $Z_{\rm RIS}= \sum_{i=1}^{N} Z_i$ using multivariate Fox's H-function. In the following Theorem, we apply a novel approach to derive the PDF $Z_i$ in terms of univariate Fox's H-function:

\begin{my_theorem}\label{th:prod_pdf_phase}
	The PDF of effective channel $Z_i$ for a single-element RIS under the combined effect of generalized asymmetric channels, user mobility, and phase errors is given as
	\begin{eqnarray}\label{eq:prod_pdf_proof_phase}
	&f_{Z_{i}}(x) = \sum_{k_{i}=0}^{\infty} \sum_{j_{i}=1}^{n} \psi_{i} x^{-1} \nonumber\\ &H_{2,5}^{3,1} \bigg[\begin{array}{c}\zeta_{i,1} \zeta_{i,2}^{} x^{} \end{array} \big\vert \begin{array}{c}(-\beta_{j_{i}},\frac{a}{2}), (1,q) \\  V_{i}, (0,q)\end{array}\bigg]
	\end{eqnarray}
	where $\psi_{i}=\psi_{i,1} \psi_{i,2} \zeta_{i,1}^{-(\mu_{i}+k_{i})}$, $V_{i}=(\beta_{i,2},\frac{1}{\alpha_{i,2}}),(\beta_{i,1},\frac{1}{\alpha_{i,1}}),(\mu_{i}+k_{i},1),(-1-\beta_{j_{i}},\frac{a}{2})$.
\end{my_theorem}

\begin{IEEEproof}
See  Appendix A.

\end{IEEEproof}
As a special case, we  can consider i.ni.d. Rayleigh fading in both the hops from source to RIS and RIS to destination by substituting $\kappa=0$ and $\mu=1$ in \eqref{eq:hi_pdf}, and  $\alpha_{1}=1$ and $\beta_{1}=2$ in  \eqref{eq:dgg_pdf_new} to simplify  PDF expressions for fading channels. Applying the similar approach for the proof of Theorem  \ref{th:prod_pdf_phase}, we can get the PDF of $Z_i$ for Rayleigh fading channel with phase noise and mobility:
\begin{eqnarray}\label{eq:Rayleigh_prod_pdf_phase}
	&f_{Z_{i}}(x) = \sum_{j_{i}=1}^{n} B_{j_{i}} x^{-1} \nonumber\\ &H_{2,4}^{2,1} \bigg[\begin{array}{c} \frac{2}{\Omega_{i,1}} d_{2}^{\frac{a}{2}}x^{} \end{array} \big\vert \begin{array}{c}(-\beta_{j_{i}},\frac{a}{2}), (1,q) \\  (2,1),(1,1),(-1-\beta_{j_{i}},\frac{a}{2}), (0,q)\end{array}\bigg]
\end{eqnarray}

Further, we can simplify \eqref{eq:Rayleigh_prod_pdf_phase}  by neglecting the mobility effect to get the PDF of $Z_i$ for Rayleigh fading channel with phase noise:
\begin{eqnarray}\label{eq:Rayleigh_phase}
	&f_{Z_{i}^{}}(x) = x^{-1} H_{1,3}^{2,1} \bigg[\begin{array}{c} \frac{2}{\Omega_{i,1}} x^{} \end{array} \big\vert \begin{array}{c}(1,q) \\  (2,1),(1,1), (0,q)\end{array}\bigg]
\end{eqnarray}

In what follows, we use the univariate Fox's H-function representation for the PDF of $Z_i$ in Theorem \ref{th:prod_pdf_phase} to develop statistical results for $Z_{\rm RIS}= \sum_{i=1}^{N} Z_i$ in  the following Lemma:
\begin{my_lemma}\label{th:sum_pdf_ris}
The PDF and CDF of the resultant RIS channel $Z_{\rm RIS}=\sum_{i=1}^{N}Z_{i}$  are given by
\begin{eqnarray}\label{eq:sum_pdf_pe_final}
	&f_{Z_{\rm RIS}}(x) = \sum_{k_{1},\cdots,k_{N}=0}^{\infty} \sum_{j_{1},\cdots,j_{N}=1}^{n} \prod_{i=1}^{N} \psi_{i} x^{-1}
	\nonumber \\ & H_{0,1:3,5;\cdots;3,5}^{0,0:3,2;\cdots;3,2} \nonumber \\ &\bigg[\begin{array}{c} \{\zeta_{i,1} \zeta_{i,2}^{} x^{}\}_{1}^{N} \end{array} \big\vert \begin{array}{c}-:\{(-\beta_{j_{i}},\frac{a}{2}),(1,q),(1,1)\}_{i=1}^{N}\\ (1:1,\cdots,1):\{V_{i},(0,q)\}_{i=1}^{N} \end{array}\bigg] 
\end{eqnarray}
\begin{eqnarray}\label{eq:sum_cdf_pe_final}
	&F_{Z_{\rm RIS}^{}}(x) = \sum_{k_{1},\cdots,k_{N}=0}^{\infty} \sum_{j_{1},\cdots,j_{N}=1}^{n} \prod_{i=1}^{N} \psi_{i} 
	\nonumber \\ &H_{0,1:3,5;\cdots;3,5}^{0,0:3,2;\cdots;3,2} \nonumber \\ &\bigg[\begin{array}{c} \{\zeta_{i,1} \zeta_{i,2}^{} x^{}\}_{1}^{N} \end{array} \big\vert \begin{array}{c}-:\{(-\beta_{j_{i}},\frac{a}{2}),(1,q),(1,1)\}_{i=1}^{N}\\ (0:1,\cdots,1):\{V_{i},(0,q)\}_{i=1}^{N} \end{array}\bigg] 
\end{eqnarray}
where $V_{i}=(\beta_{i,2},\frac{1}{\alpha_{i,2}}),(\beta_{i,1},\frac{1}{\alpha_{i,1}}),(\mu_{i}+k_{i},1),(-1-\beta_{j_{i}},\frac{a}{2})$.
\end{my_lemma}
\begin{IEEEproof} 
See Appendix B.\end{IEEEproof}

Similarly, we can use \eqref{eq:Rayleigh_prod_pdf_phase} in Lemma \ref{th:sum_pdf_ris} to develop the PDF and CDF of  $Z_{\rm RIS}$ for Rayleigh fading with user mobility and phase noise:
\begin{eqnarray}\label{eq:Rayleigh_sum_pdf_pe_final}
	&f_{Z_{\rm RIS}^{}}(x) = \sum_{j_{1},\cdots,j_{N}=1}^{n} \prod_{i=1}^{N} B_{j_{i}} x^{-1}
	H_{0,1:3,4;\cdots;3,4}^{0,0:2,2;\cdots;2,2} \nonumber \\ &\bigg[\begin{array}{c} \{ \frac{2}{\Omega_{i,1}} d_{2}^{\frac{a}{2}}x^{}\}_{1}^{N} \end{array} \big\vert \begin{array}{c}-:\{(-\beta_{j_{i}},\frac{a}{2}),(1,q),(1,1)\}_{i=1}^{N}\\ (1:1,\cdots,1):\{V_{i},(0,q)\}_{i=1}^{N} \end{array}\bigg] 
\end{eqnarray}
\begin{eqnarray}\label{eq:Rayleigh_sum_cdf_pe_final}
	&F_{Z_{\rm RIS}^{}}(x) = \sum_{j_{1},\cdots,j_{N}=1}^{n} \prod_{i=1}^{N} B_{j_{i}} 
	H_{0,1:3,4;\cdots;3,4}^{0,0:2,2;\cdots;2,2} \nonumber \\ &\bigg[\begin{array}{c} \{ \frac{2}{\Omega_{i,1}} d_{2}^{\frac{a}{2}}x^{}\}_{1}^{N} \end{array} \big\vert \begin{array}{c}-:\{(-\beta_{j_{i}},\frac{a}{2}),(1,q),(1,1)\}_{i=1}^{N}\\ (0:1,\cdots,1):\{V_{i},(0,q)\}_{i=1}^{N} \end{array}\bigg] 
\end{eqnarray}
where $V_{i}=(2,1),(1,1),(-1-\beta_{j_{i}},\frac{a}{2})$.

Further, we can use \eqref{eq:Rayleigh_phase} in Lemma \ref{th:sum_pdf_ris} to represent the PDF and CDF of  $Z_{\rm RIS}$ for Rayleigh fading channels with phase noise as
\begin{eqnarray}\label{eq:Rayleigh_sum_pdf_phase}
	&f_{Z_{\rm RIS}^{}}(x) = x^{-1}
	H_{0,1:2,3;\cdots;2,3}^{0,0:2,1;\cdots;2,1} \nonumber \\ &\hspace{-4mm}\bigg[\begin{array}{c} \{ \frac{2}{\Omega_{i,1}} x^{}\}_{1}^{N} \end{array} \big\vert \begin{array}{c}-:\{(1,q),(1,1)\}_{i=1}^{N}\\ (1:1,\cdots,1):\{(2,1),(1,1),(0,q)\}_{i=1}^{N} \end{array}\bigg] 
\end{eqnarray}
\begin{eqnarray}\label{eq:Rayleigh_sum_cdf_phase}
	&F_{Z_{\rm RIS}^{}}(x) =  
	H_{0,1:2,3;\cdots;2,3}^{0,0:2,1;\cdots;2,1} \nonumber \\ &\hspace{-4mm}\bigg[\begin{array}{c} \{ \frac{2}{\Omega_{i,1}} x^{}\}_{1}^{N} \end{array} \big\vert \begin{array}{c}-:\{(1,q),(1,1)\}_{i=1}^{N}\\ (0:1,\cdots,1):\{(2,1),(1,1),(0,q)\}_{i=1}^{N} \end{array}\bigg] 
\end{eqnarray}
where $V_{i}=(2,1),(1,1),(-1-\beta_{j_{i}},\frac{a}{2})$.

Although computational routines \cite{Alhennawi2016} are available for multivariate Fox's H-function in Lemma \ref{th:sum_pdf_ris}, it is desirable to simplify the analytical expressions using univariate Fox's H-function for better insights into the system performance. 

Using the inequality of  arithmetic mean and geometric mean, we can express $\sum_{i=1}^{N}\frac{Z_{i}}{N} \geq  \prod_{i=1}^{N}Z_{i}^{\frac{1}{N}}$. Thus, an upper bound on the CDF of $Z_{\rm RIS}$ can be expressed as 
\begin{equation}\label{eq:cdf_approximation}
F_{Z_{\rm RIS}^{}}(x) \leq F_{Y_{\rm RIS}^{}}\bigg((\frac{x}{N})^{N}\bigg)  
\end{equation} 
where $Y_{\rm RIS}=\prod_{i=1}^{N}Z_{i}$ is the product of $Z_i$ with $i=1, 2, 3, \cdots N$. 
Similarly, an upper bound on the PDF of $Z_{\rm RIS}$ is:
\begin{equation}\label{eq:pdf_approximation}
 f_{Z_{\rm RIS}}(x) \leq (\frac{x}{N})^{N-1} f_{Y_{\rm RIS}^{}}\big((\frac{x}{N})^{N}\big)
 \end{equation}
 It can be seen from \eqref{eq:cdf_approximation} and \eqref{eq:pdf_approximation}  that statistical results of $Z_{\rm RIS}$ requires the PDF and CDF of $N$-product of $Z_i$. We can use the Mellin inverse transform to get the PDF of  $Y_{\rm RIS}=\prod_{i=1}^{N}Z_{i}$  \cite{chapala2021unified,chapala2022vtcmultiris}:
  \begin{eqnarray}\label{eq:prod_pdf}
 &f_{Y_{\rm RIS}^{}}(x) = \frac{x^{-1}}{2\pi \J} \int\limits_{\mathcal{L}} \Aver{Y_{\rm RIS}^{r}} x^{-r} dr \nonumber \\&= \frac{x^{-1}}{2\pi \J} \int\limits_{\mathcal{L}} \prod_{i=1}^{N} \Aver{(Z_{i}^{})^{r}} x^{-r}dr
 \end{eqnarray}
 where $\Aver{\cdot}$ denotes the expectation operator.
\begin{my_lemma}
	Upper bounds on the PDF and CDF of the resultant RIS channel $Z_{\rm RIS}$ are given by
	\begin{eqnarray}\label{eq:pdf_ris_approx}
		&f_{Z_{\rm RIS}^{}}(x) \leq (\frac{x}{N})^{-1} \sum_{k_{1},\cdots,k_{N}=0}^{\infty} \sum_{j_{1},\cdots,j_{N}=1}^{n} \prod_{i=1}^{N} \psi_{i} \nonumber \\ &\hspace{-4mm}H_{N+1,4N+1}^{3N,N} \bigg[\begin{array}{c}\prod_{i=1}^{N}\zeta_{i,1} \zeta_{i,2}^{} (\frac{x}{N})^{N} \end{array} \big\vert \begin{array}{c}\{(-\beta_{j_{i}},\frac{a}{2})\}_{1}^{N}, (1,q) \\ \{V_{i}\}_{1}^{N}, (0,q) \end{array}\bigg]
	\end{eqnarray}
		\begin{eqnarray}\label{eq:cdf_ris_approx}
			&F_{Z_{\rm RIS}^{}}(x) \leq  \sum_{k_{1},\cdots,k_{N}=0}^{\infty} \sum_{j_{1},\cdots,j_{N}=1}^{n} \prod_{i=1}^{N} \psi_{i} H_{N+2,4N+2}^{3N,N+1} \nonumber \\ &\hspace{-2mm}\bigg[\begin{array}{c}\prod_{i=1}^{N}\zeta_{i,1} \zeta_{i,2}^{} (\frac{x}{N})^{N} \end{array} \big\vert \begin{array}{c}\{(-\beta_{j_{i}},\frac{a}{2})\}_{1}^{N},(1,1), (1,q) \\ \{V_{i}\}_{1}^{N},(0,1), (0,q) \end{array}\bigg]
		\end{eqnarray}
	where $V_{i}=(\beta_{i,2},\frac{1}{\alpha_{i,2}}),(\beta_{i,1},\frac{1}{\alpha_{i,1}}),(\mu_{i}+k_{i},1),(-1-\beta_{j_{i}},\frac{a}{2})$.
\end{my_lemma}
\begin{IEEEproof}
		See Appendix C.
\end{IEEEproof}
As a special case, we can use the proof of Appendix C with \eqref{eq:Rayleigh_prod_pdf_phase} for Rayleigh fading channels with phase noise and mobility to get  upper bounds on the PDF and CDF of $Z_{\rm RIS}$:
\begin{eqnarray}\label{eq:Rayleigh_pdf_ris_approx}
	&f_{Z_{\rm RIS}}(x) \leq (\frac{x}{N})^{-1} \sum_{j_{1},\cdots,j_{N}=1}^{n} \prod_{i=1}^{N} B_{j_{i}} H_{N+1,3N+1}^{2N,N} \nonumber \\ &\bigg[\begin{array}{c} \prod_{i=1}^{N}\frac{2}{\Omega_{i,1}} d_{2}^{\frac{a}{2}} (\frac{x}{N})^{N} \end{array} \big\vert \begin{array}{c}\{(-\beta_{j_{i}},\frac{a}{2})\}_{1}^{N}, (1,q) \\ \{V_{i}\}_{1}^{N}, (0,q) \end{array}\bigg]
\end{eqnarray}
\begin{eqnarray}\label{eq:Rayleigh_cdf_ris_approx}
	&F_{Z_{\rm RIS}^{}}(x) \leq\sum_{j_{1},\cdots,j_{N}=1}^{n} \prod_{i=1}^{N} B_{j_{i}} H_{N+2,3N+2}^{2N,N+1} \nonumber \\ &\bigg[\begin{array}{c}\prod_{i=1}^{N} \frac{2}{\Omega_{i,1}} d_{2}^{\frac{a}{2}} (\frac{x}{N})^{N} \end{array} \big\vert \begin{array}{c}\{(-\beta_{j_{i}},\frac{a}{2})\}_{1}^{N},(1,1), (1,q) \\ \{V_{i}\}_{1}^{N},(0,1), (0,q) \end{array}\bigg]
\end{eqnarray}
where $V_{i}=(2,1),(1,1),(-1-\beta_{j_{i}},\frac{a}{2})$.
Similarly, upper bounds for the PDF and CDF of $Z_{\rm RIS}$ for Rayleigh fading channels with phase noise can be derived as
\begin{eqnarray}\label{eq:Rayleigh_pdf_ris_approx_phase}
	&f_{Z_{\rm RIS}^{}}(x) \leq (\frac{x}{N})^{-1} H_{N+1,3N+1}^{2N,N} \nonumber \\ &\bigg[\begin{array}{c}\prod_{i=1}^{N} \frac{2}{\Omega_{i,1}} (\frac{x}{N})^{N} \end{array} \big\vert \begin{array}{c}(1,q) \\ \{(2,1),(1,1)\}_{1}^{N}, (0,q) \end{array}\bigg]
\end{eqnarray}

\begin{eqnarray}\label{eq:Rayleigh_cdf_ris_approx_phase}
	&F_{Z_{\rm RIS}^{}}(x) \leq  H_{N+2,3N+2}^{2N,N+1} \nonumber \\ &\bigg[\begin{array}{c}\prod_{i=1}^{N} \frac{2}{\Omega_{i,1}} (\frac{x}{N})^{N} \end{array} \big\vert \begin{array}{c}(1,1), (1,q) \\ \{(2,1),(1,1)\}_{1}^{N},(0,1), (0,q) \end{array}\bigg]
\end{eqnarray}
where $V_{i}=(2,1),(1,1),(-1-\beta_{j_{i}},\frac{a}{2})$.
\subsection{With Direct Link ($\omega=1$)}
In this subsection, we provide statistical results for the considered RIS-assisted system with direct link by applying the maximal ratio combining (MRC) receiver at the destination for the received signals from the RIS and direct transmissions. Here, we develop statistical results  for the resultant SNR of the RIS-assisted system given as $\gamma_{\rm RISD}=\gamma_{\rm RIS}+\gamma_{d} = \bar{\gamma}_{\rm RIS} Z_{\rm RIS}^{2} + \bar{\gamma}_{d} Z_{d}^{2}$,	where  $\bar{\gamma}_{\rm RIS}=\frac{H_{l}^{2}P_{t}}{\sigma_{v}^{2}}$ and $\bar{\gamma}_{d}=\frac{P_{t}}{\sigma_{v}^{2}}$.

The PDF and CDF of SNR of RIS link $f_{\gamma_{\rm RIS}^{}}(\gamma)$ can be expressed using mathematical transformation as
\begin{eqnarray}\label{eq_ref_zaf}
	&f_{\gamma_{\rm RIS}^{}}(\gamma)=\frac{1}{2\sqrt{\bar{\gamma}_{\rm RIS}\gamma}}f_{Z_{\rm RIS}^{}}(\sqrt{\frac{\gamma}{\bar{\gamma}_{\rm RIS}}})  \\
	&F_{\gamma_{\rm RIS}^{}}(\gamma)=F_{Z_{\rm RIS}^{}}(\sqrt{\frac{\gamma}{\bar{\gamma}_{\rm RIS}}})   
\end{eqnarray} 

	We denote the direct link path gain component as $h^{\rm rwp}_{l,d}=r^{-\frac{a}{2}}$, where $r$ is random due to the mobility of user. Thus, the PDF of short-term fading of  the direct link combined with mobility model can be expressed as
\begin{eqnarray}\label{eq:comb_pdf_dl}
	f_{Z_{d}}(x) = \int_{0}^{d_{}} f_{Z_{d}^{}}(x/r) f_{r}(r) dr
\end{eqnarray}
Applying standard mathematical procedure, the  PDF of the SNR $\bar{\gamma}_{d} Z_{d}^{2}$ for the  direct-link channel combined with mobility model   is given as
\begin{eqnarray}\label{eq:comb_genk_mob}
	&f_{\gamma_{d}}(x) = \sum_{j=1}^{n} \psi_{d} x^{M+m-1} H_{0,3}^{2,1} \nonumber \\ &\hspace{-6mm}\bigg[\begin{array}{c} \frac{b  d^{\frac{a}{2}}x}{2} \end{array} \big\vert \begin{array}{c} (\beta_{j}+\frac{a(M+m)}{2},\frac{a}{2}) \\ (\frac{M-m}{2},\frac{1}{2}),(\frac{m-M}{2},\frac{1}{2}),(\beta_{j}+\frac{a(M+m)}{2}+1,\frac{a}{2})\end{array}\bigg]
\end{eqnarray}
where $\psi_{d}=B_{j} \frac{b^{M+m}d_{}^{\frac{a(M+m)}{2}}}{2^{M+m} \Gamma(m)\Gamma(M)}$.

To derive the PDF of $\gamma_{\rm RISD}= \gamma_{\rm RIS}+\gamma_{d}$, we first compute the MGF of $\gamma_{\rm RIS}$ and $\gamma_{d}$ and apply the inverse Laplace transform, as presented in the following Theorem:
\begin{my_theorem}
	Exact  and upper bounds for the  PDF and CDF of the resultant SNR $\gamma_{\rm RISD}=\gamma_{\rm RIS}+\gamma_{d}$ are given in \eqref{eq:pdf_snr_final} and \eqref{eq:cdf_snr_final}.
		\begin{figure*}
		\begin{flalign}\label{eq:pdf_snr_final}
			f_{\gamma_{\rm RISD^{}}}(\gamma)& =\frac{1}{4\gamma} \sum_{k_{1},\cdots,k_{N}=0}^{\infty} \sum_{j_{1},\cdots,j_{N}=1}^{n} \sum_{j=1}^{n}  \prod_{i=1}^{N} B_{j} \psi_{i} \frac{1}{\Gamma(m)\Gamma(M)}\nonumber \\ 
			&H_{1,2:3,5;\cdots;3,5;2,3}^{0,1:3,2;\cdots;3,2;2,2} \bigg[\begin{array}{c} \{\zeta_{i,1} \zeta_{i,2}^{} \sqrt{\frac{\gamma}{\bar{\gamma}_{\rm RIS}}}^{}\}_{1}^{N} \\ \frac{b  d^{\frac{a}{2}}}{2} \sqrt{\frac{\gamma}{\bar{\gamma}_{d}}}^{}\end{array} \big\vert \begin{array}{c}(1:\frac{1}{2},\cdots,\frac{1}{2},0):V_{1}\\ (1:1,\cdots,1,0), (1:\frac{1}{2},\cdots,\frac{1}{2},\frac{1}{2}):V_{2} \end{array}\bigg] \nonumber \\  
			&\leq \frac{N}{4\gamma} \sum_{k_{1},\cdots,k_{N}=0}^{\infty} \sum_{j_{1},\cdots,j_{N}=1}^{n} \sum_{j=1}^{n}   \prod_{i=1}^{N} \psi_{i} \frac{B_{j}}{\Gamma(m)\Gamma(M)}\nonumber \\ &H_{0,1:N+1,4N+1;2,3}^{0,0:3N,N;2,2} \bigg[\begin{array}{c}\prod_{i=1}^{N}\zeta_{i,1} \zeta_{i,2}^{} (\frac{1}{N}\sqrt{\frac{\gamma}{\bar{\gamma}_{\rm RIS}}})^{N} \\ \frac{b  d^{\frac{a}{2}}}{2} \sqrt{\frac{\gamma}{\bar{\gamma}_{d}}}^{}\end{array} \big\vert \begin{array}{c}-:\{(-\beta_{j_{i}},\frac{a}{2})\}_{1}^{N}, (1,q);(\beta_{j}+a(M+m),\frac{a}{2}),(1,\frac{1}{2}) \\ (1:\frac{N}{2},\frac{1}{2}):V_{3}, (0,q);(M,\frac{1}{2}),(m,\frac{1}{2}),(1+\beta_{j}+a(M+m),\frac{a}{2}) \end{array}\bigg]
		\end{flalign}
	
		\begin{flalign}\label{eq:cdf_snr_final}
			F_{\gamma_{\rm RISD^{}}}(\gamma) &= \frac{1}{4} \sum_{k_{1},\cdots,k_{N}=0}^{\infty} \sum_{j_{1},\cdots,j_{N}=1}^{n} \sum_{j=1}^{n}  \prod_{i=1}^{N} B_{j} \psi_{i} \frac{1}{\Gamma(m)\Gamma(M)}\nonumber \\ 
			&H_{1,2:3,5;\cdots;3,5;2,3}^{0,1:3,2;\cdots;3,2;2,2} \bigg[\begin{array}{c} \{\zeta_{i,1} \zeta_{i,2}^{} \sqrt{\frac{\gamma}{\bar{\gamma}_{\rm RIS}}}^{}\}_{1}^{N} \\ \frac{b  d^{\frac{a}{2}}}{2} \sqrt{\frac{\gamma}{\bar{\gamma}_{d}}}^{}\end{array} \big\vert \begin{array}{c}(1:\frac{1}{2},\cdots,\frac{1}{2},0):V_{1}\\ (1:1,\cdots,1,0), (0:\frac{1}{2},\cdots,\frac{1}{2},\frac{1}{2}):V_{2} \end{array}\bigg] \nonumber \\
		&\leq	\frac{N}{4} \sum_{k_{1},\cdots,k_{N}=0}^{\infty} \sum_{j_{1},\cdots,j_{N}=1}^{n} \sum_{j=1}^{n}   \prod_{i=1}^{N} \psi_{i} \frac{B_{j}}{\Gamma(m)\Gamma(M)}\nonumber \\ &H_{0,1:N+1,4N+1;2,3}^{0,0:3N,N;2,2} \bigg[\begin{array}{c}\prod_{i=1}^{N}\zeta_{i,1} \zeta_{i,2}^{} (\frac{1}{N}\sqrt{\frac{\gamma}{\bar{\gamma}_{\rm RIS}}})^{N} \\ \frac{b  d^{\frac{a}{2}}}{2} \sqrt{\frac{\gamma}{\bar{\gamma}_{d}}}^{}\end{array} \big\vert \begin{array}{c}-:\{(-\beta_{j_{i}},\frac{a}{2})\}_{1}^{N}, (1,q);(\beta_{j}+a(M+m),\frac{a}{2}),(1,\frac{1}{2}) \\ (0:\frac{N}{2},\frac{1}{2}):V_{3}, (0,q);(M,\frac{1}{2}),(m,\frac{1}{2}),(1+\beta_{j}+a(M+m),\frac{a}{2}) \end{array}\bigg]
		\end{flalign}
		where $V_{1}=\{(-\beta_{j_{i}},\frac{a}{2}),(1,q),(1,1)\}_{i=1}^{N};(\beta_{j}+a(M+m),\frac{a}{2}),(1,\frac{1}{2})$, $V_{2}=\{(\beta_{i,2},\frac{1}{\alpha_{i,2}}),(\beta_{i,1},\frac{1}{\alpha_{i,1}}),(\mu_{i}+k_{i},1),(-1-\beta_{j_{i}},\frac{a}{2}),(0,q)\}_{i=1}^{N};(M,\frac{1}{2}),(m,\frac{1}{2}),(1+\beta_{j}+a(M+m),\frac{a}{2})$, and $V_{3}=\{(\beta_{i,2},\frac{1}{\alpha_{i,2}}),(\beta_{i,1},\frac{1}{\alpha_{i,1}}),(\mu_{i}+k_{i},1),(-1-\beta_{j_{i}},\frac{a}{2})\}_{1}^{N}$.
		\hrule
	\end{figure*}
\end{my_theorem}
\begin{IEEEproof}
	Refer to Appendix D.
\end{IEEEproof}

In the following section, we use the derived statistical results to analysis the performance of RIS-assisted system with/without direct link under the combined effect of phase noise and mobility.

\section{Performance Analysis}
     In this section we analyze the outage probability and average BER of RIS system without and with direct link. We derive exact expressions, upper bound and asymptotic expressions at high SNR for the  considered systems. 
      
 To analyze the system performance, we use the CDF of the SNR presented in the previous section.  For the case without direct link $\omega =0$, the CDF of the SNR $\gamma_{\rm RIS} $ is	$F_{\gamma_{\rm RIS}}(\gamma) = F_{Z_{\rm RIS}}(\sqrt{\frac{\gamma}{\bar{\gamma}_{\rm RIS}}})$, where     $F_{Z_{\rm RIS}}(x)$ is derived in Section III-A for various scenarios. Similarly, the CDF of the SNR with direct link (i.e., $\omega =1$) $F_{\gamma_{\rm RISD}}(\gamma)$  is derived in Section III-B.
     
\subsection{Outage Probability}\label{sec:op}
Outage probability is a performance metric to characterize the impact of fading in a communication system. Mathematically, it can be defined as  the probability of SNR falling below a threshold value $\gamma_{\rm th}$ i.e.,  $ P_{\rm out}=Pr(\gamma \le \gamma_{\rm th})$.  Thus, an exact expression for the RIS system  can be expressed as $F_{\gamma_{\rm RIS}}(\gamma_{\rm th})$ for without direct link and $F_{\gamma_{\rm RISD}}(\gamma_{\rm th})$ with direct link. 

To get the outage probability at a high SNR, we can apply the asymptotic series expansion \cite{Kilbas_FoxH,Rahama2018} on the CDF functions represented in terms of $N$-multivariate (exact expression  without direct link \eqref{eq:sum_cdf_pe_final}), $N+1$-multivariate (exact expression  with direct link \eqref{eq:cdf_snr_final}), univariate (upper bound  without direct link \eqref{eq:cdf_ris_approx}), bivariate (upper bound  with direct link \eqref{eq:cdf_snr_final}) Fox's H-function.

To illustrate, we  apply \cite[eq. (30)]{Rahama2018} on the $N+1$-multivariate Fox's H-function in \eqref{eq:cdf_snr_final}, to represent the outage probability  asymptotically at a high SNR regime for the RIS-assisted system with direct link: 
 \begin{eqnarray}\label{eq:outage_asymp}
 	&P_{{\rm out, \rm RISD}}^{\infty} = \frac{1}{4} \sum_{k_{1},\cdots,k_{N}=0}^{\infty} \sum_{j_{1},\cdots,j_{N}=1}^{n} \sum_{j=1}^{n}  \prod_{i=1}^{N} B_{j} \psi_{i} \nonumber \\ 
 	&\frac{1}{\Gamma(m)\Gamma(M)}\frac{\Gamma(\frac{p_{1}}{2}+\cdots+\frac{p_{N}}{2})}{\Gamma(p_{1}+\cdots+p_{N}) \Gamma(1+\frac{p_{1}}{2}+\cdots+\frac{p_{N}}{2}+\frac{p_{N+1}}{2})} \nonumber\\ &  \prod_{i=1}^{N} \frac{\Gamma(\beta_{i,2}-\frac{p_{i}}{\alpha_{i,2}}) \Gamma(\beta_{i,1}-\frac{p_{i}}{\alpha_{i,1}}) \Gamma(\mu_{i}+k_{i}-p_{i}) \Gamma(1+\beta_{j_{i}}+\frac{a p_{i}}{2}) \Gamma(p_{i})}{\Gamma(2+\beta_{j_{i}}+\frac{a p_{i}}{2})\Gamma(1-q p_{i})\Gamma(1+q p_{i})} \nonumber \\ & (\zeta_{i,1} \zeta_{i,2}^{} \sqrt{\frac{\gamma_{\rm th}}{\bar{\gamma}_{\rm RIS}}}^{})^{p_{i}-1} \nonumber \\ &\frac{\Gamma(M-\frac{p_{N+1}}{2}) \Gamma(m-\frac{p_{N+1}}{2}) \Gamma(1-\beta_{j}+a(M+m)+\frac{a p_{N+1}}{2}) \Gamma(\frac{p_{N+1}}{2})}{\Gamma(\beta_{j}+a(M+m)+\frac{a p_{N+1}}{2})} \nonumber \\ &(\frac{b  d^{\frac{a}{2}}}{2} \sqrt{\frac{\gamma_{\rm th}}{\bar{\gamma}_{d}}}^{})^{p_{N+1}-1}
 \end{eqnarray}
 where $p_{i} = \min\{\alpha_{i,1}\beta_{i,1},\alpha_{i,2}\beta_{i,2},\mu_{i}\}$ for $i\in(1, N)$, $p_{N+1}=\min\{2M,2m\}$ is the dominant term. 
 
 To get the outage diversity order of the system, we can express \eqref{eq:outage_asymp} as $P_{\rm out, RISD}^{\infty} \propto \bar{\gamma}^{-G_{\rm out}}$ resulting $G_{\rm out}= \sum_{i}^{N+1}\frac{p_{i}}{2}$. Thus, the diversity order provides various design criteria depending on the fading and system parameters. It can be seen that diversity order is independent of the parameter $\kappa$.

\subsection{ Average BER}\label{sec:ber}
Average BER is used to quantify the reliability of data transmissions. For binary modulations, the average BER using the CDF of SNR is given as \cite{Ansari2011_ber}
\begin{equation}\label{eq:gen_ber}
	\bar{P}_{e} =  \frac{q^{p}}{2\Gamma(p)} \int_{0}^{\infty} \gamma^{p-1} e^{-q\gamma} F_{\gamma} (\gamma) d\gamma
\end{equation}
where $p$ and $q$ are modulation specific parameters. 

For the RIS-assisted system with direct link ($\omega=1$), we substitute the exact expression of the CDF (see \eqref{eq:cdf_snr_final}) in \eqref{eq:gen_ber} and expand the definition of multivariate Fox-H function and interchange the order of integration to get
\begin{eqnarray}\label{eq:ber_1}
	&\bar{P}_{e, {\rm RISD}} = \frac{q^{p}}{2\Gamma(p)} \sum_{k_{1},\cdots,k_{N}=0}^{\infty} \sum_{j_{1},\cdots,j_{N}=1}^{n} \sum_{j=1}^{n}  \prod_{i=1}^{N}  \nonumber \\ &B_{j} \frac{\psi_{i}}{4}\frac{1}{\Gamma(m)\Gamma(M)} \frac{1}{2\pi \J} \int_{\mathcal{L}_{i}} (\zeta_{i,2}^{} \zeta_{i,1} \sqrt{\frac{1}{\bar{\gamma}_{\rm RIS}}}^{})^{s_{i}} \nonumber \\ &\frac{\Gamma(1+\beta_{j_{i}}+\frac{a}{2}s_{i}) \Gamma(\beta_{i,1}-\frac{s_{i}}{\alpha_{i,1}}) \Gamma(\beta_{i,2}-\frac{s_{i}}{\alpha_{i,2}})  \Gamma(\mu_{i}+k_{i}-s_{i})  \Gamma(s_{i})}{\Gamma(2+\beta_{j_{i}}+\frac{a}{2}s_{i})\Gamma(1-q s_{i})\Gamma(1+q s_{i})} \nonumber \\
	& \frac{\Gamma(\sum_{i=1}^{N}\frac{s_{i}}{2})}{\Gamma(\sum_{i}^{N}s_{i})}   \frac{1}{2\pi \J} \int_{\mathcal_{L}_{d}} (\frac{b  d^{\frac{a}{2}}}{2} \sqrt{\frac{1}{\bar{\gamma}_{d}}}^{})^{s_{d}} \Gamma(\frac{s_{d}}{2})\nonumber \\ &\frac{\Gamma(M-\frac{s_{d}}{2})\Gamma(m-\frac{s_{d}}{2}) \Gamma(1-\beta_{j}-a(M+m)+\frac{a}{2}s_{d})}{\Gamma(-\beta_{j}-a(M+m)+\frac{a}{2}s_{d})\Gamma(\sum_{i=1}^{N}\frac{s_{i}}{2}+\frac{s_{d}}{2}+1)}   \nonumber \\ &  \Big(\int_{0}^{\infty} \gamma^{p-1} e^{-q\gamma} \gamma^{\sum_{i=1}^{N}\frac{s_{i}}{2}+\frac{s_{d}}{2}} d\gamma\Big) ds_{i} ds_{d} 
\end{eqnarray}

\begin{figure*}
	\begin{flalign}\label{eq:ber}
		\bar{P}_{e, {\rm RISD}}& = \frac{1}{8\Gamma(p)} \sum_{k_{1},\cdots,k_{N}=0}^{\infty} \sum_{j_{1},\cdots,j_{N}=1}^{n} \sum_{j=1}^{n}  \prod_{i=1}^{N} B_{j} \psi_{i} \frac{1}{\Gamma(m)\Gamma(M)}\nonumber \\ 
		&H_{2,2:3,5;\cdots;3,5;2,3}^{0,2:3,2;\cdots;3,2;2,2} \bigg[\begin{array}{c} \{\zeta_{i,1} \zeta_{i,2}^{} \sqrt{\frac{1}{q\bar{\gamma}_{\rm RIS}}}^{}\}_{1}^{N} \\ \frac{b  d^{\frac{a}{2}}}{2} \sqrt{\frac{1}{q\bar{\gamma}_{d}}}^{}\end{array} \big\vert \begin{array}{c}(1:\frac{1}{2},\cdots,\frac{1}{2},0),(1-p:\frac{1}{2},\cdots,\frac{1}{2},\frac{1}{2}):V_{1}\\ (1:1,\cdots,1,0), (0:\frac{1}{2},\cdots,\frac{1}{2},\frac{1}{2}):V_{2} \end{array}\bigg]\\
		&\leq \frac{N}{8\Gamma(p)} \sum_{k_{1},\cdots,k_{N}=0}^{\infty} \sum_{j_{1},\cdots,j_{N}=1}^{n} \sum_{j=1}^{n}   \prod_{i=1}^{N} \psi_{i} \frac{B_{j}}{\Gamma(m)\Gamma(M)}\nonumber \\ &H_{1,1:N+1,4N+1;2,3}^{0,1:3N,N;2,2} \bigg[\begin{array}{c}\prod_{i=1}^{N}\zeta_{i,1} \zeta_{i,2}^{} (\frac{1}{N}\sqrt{\frac{1}{q\bar{\gamma}_{\rm RIS}}})^{N} \\ \frac{b  d^{\frac{a}{2}}}{2} \sqrt{\frac{1}{q\bar{\gamma}_{d}}}^{}\end{array} \big\vert \begin{array}{c}(1-p:\frac{N}{2},\frac{1}{2}): V_{3}\\ (0:\frac{N}{2},\frac{1}{2}):V_{4} \end{array}\bigg]\label{eq:ber_approx}
	\end{flalign}
	where $V_{1}=\{(-\beta_{j_{i}},\frac{a}{2}),(1,q),(1,1)\}_{i=1}^{N};(\beta_{j}+a(M+m),\frac{a}{2}),(1,\frac{1}{2})$, $V_{2}=\{(\beta_{i,2},\frac{1}{\alpha_{i,2}}),(\beta_{i,1},\frac{1}{\alpha_{i,1}}),(\mu_{i}+k_{i},1),(-1-\beta_{j_{i}},\frac{a}{2}),(0,q)\}_{i=1}^{N};(M,\frac{1}{2}),(m,\frac{1}{2}),(1+\beta_{j}+a(M+m),\frac{a}{2})$, $V_{3}=\{(-\beta_{j_{i}},\frac{a}{2})\}_{1}^{N}, (1,q);(\beta_{j}+a(M+m),\frac{a}{2}),(1,\frac{1}{2})$,  and $V_{4}=\{(\beta_{i,2},\frac{1}{\alpha_{i,2}}),(\beta_{i,1},\frac{1}{\alpha_{i,1}}),(\mu_{i}+k_{i},1),(-1-\beta_{j_{i}},\frac{a}{2})\}_{1}^{N}, (0,q);(M,\frac{1}{2}),(m,\frac{1}{2}),(1+\beta_{j}+a(M+m),\frac{a}{2})$.
	\hrule
\end{figure*}

Now, the inner integral is solved as $I=\int_{0}^{\infty} \gamma^{p-1} e^{-q\gamma} \gamma^{\sum_{i=1}^{N}\frac{s_{i}}{2}+\frac{s_{d}}{2}} d\gamma=(\frac{1}{q})^{p+\sum_{i=1}^{N}\frac{s_{i}}{2}+\frac{s_{d}}{2}} \Gamma(p+\sum_{i=1}^{N}\frac{s_{i}}{2}+\frac{s_{d}}{2})$. We substitute it back and apply the definition of multivariate Fox's H-function to get an exact expression of the average BER \eqref{eq:ber}. Using the upper bound of the CDF (see \eqref{eq:cdf_snr_final}) in \eqref{eq:gen_ber}, apply the definition of bi-variate Fox's H-function to get \eqref{eq:ber_approx}. 

Similar to the direct link scenario, we can develop results for without direct link, as given by \eqref{eq:ris_ber} and \eqref{eq:ris_ber_approx}, respectively.
\begin{figure*}
	\begin{eqnarray}\label{eq:ris_ber}
		&\bar{P}_{e,\rm {RIS}} = \frac{1}{8\Gamma(p)} \sum_{k_{1},\cdots,k_{N}=0}^{\infty} \sum_{j_{1},\cdots,j_{N}=1}^{n} \prod_{i=1}^{N} \psi_{i} \nonumber \\ 
		&H_{2,2:3,5;\cdots;3,5}^{0,2:3,2;\cdots;3,2} \bigg[\begin{array}{c} \{\zeta_{i,1} \zeta_{i,2}^{} \sqrt{\frac{1}{q\bar{\gamma}_{\rm RIS}}}^{}\}_{1}^{N} \end{array} \big\vert \begin{array}{c}(1:\frac{1}{2},\cdots,\frac{1}{2}),(1-p:\frac{1}{2},\cdots,\frac{1}{2}):V_{1}\\ (1:1,\cdots,1), (0:\frac{1}{2},\cdots,\frac{1}{2}):V_{2} \end{array}\bigg]\\
		&\leq \frac{N}{8\Gamma(p)} \sum_{k_{1},\cdots,k_{N}=0}^{\infty} \sum_{j_{1},\cdots,j_{N}=1}^{n} \prod_{i=1}^{N} \psi_{i} \nonumber \\ &H_{1,1:N+1,4N+1}^{0,1:3N,N} \bigg[\begin{array}{c}\prod_{i=1}^{N}\zeta_{i,1} \zeta_{i,2}^{} (\frac{1}{N}\sqrt{\frac{1}{q\bar{\gamma}_{\rm RIS}}})^{N} \end{array} \big\vert \begin{array}{c}(1-p:\frac{N}{2}): V_{3}\\ (0:\frac{N}{2}):V_{4} \end{array}\bigg]\label{eq:ris_ber_approx}
	\end{eqnarray}
	where $V_{1}=\{(-\beta_{j_{i}},\frac{a}{2}),(1,q),(1,1)\}_{i=1}^{N}$, $V_{2}=\{(\beta_{i,2},\frac{1}{\alpha_{i,2}}),(\beta_{i,1},\frac{1}{\alpha_{i,1}}),(\mu_{i}+k_{i},1),(-1-\beta_{j_{i}},\frac{a}{2}),(0,q)\}_{i=1}^{N}$, $V_{3}=\{\{(-\beta_{j_{i}},\frac{a}{2})\}_{1}^{N}, (1,q)\}$, and $V_{4}=\{\{(\beta_{i,2},\frac{1}{\alpha_{i,2}}),(\beta_{i,1},\frac{1}{\alpha_{i,1}}),(\mu_{i}+k_{i},1),(-1-\beta_{j_{i}},\frac{a}{2})\}_{1}^{N}, (0,q)\}$.
	\hrule
\end{figure*}

\begin{figure}
	\centering
	{\includegraphics[scale=0.45]{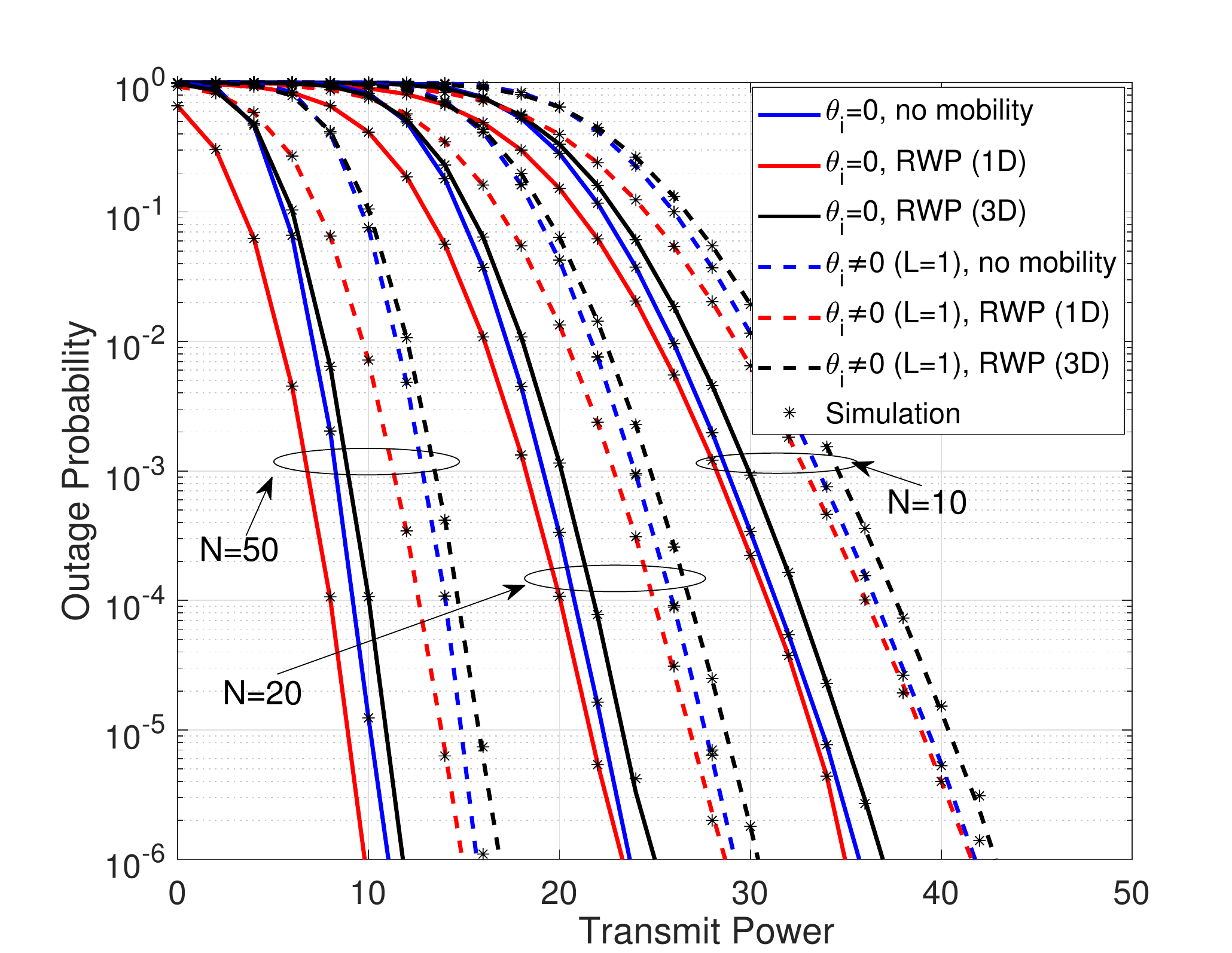}} 
	\caption{Outage performance of  the RIS-assisted system  without direct link ($\omega=0$) for    phase noise level $L=1$ at different mobility parameters.}
	\label{fig:outage_ris_mob}
\end{figure}

\begin{figure}
	\centering
	{\includegraphics[scale=0.45]{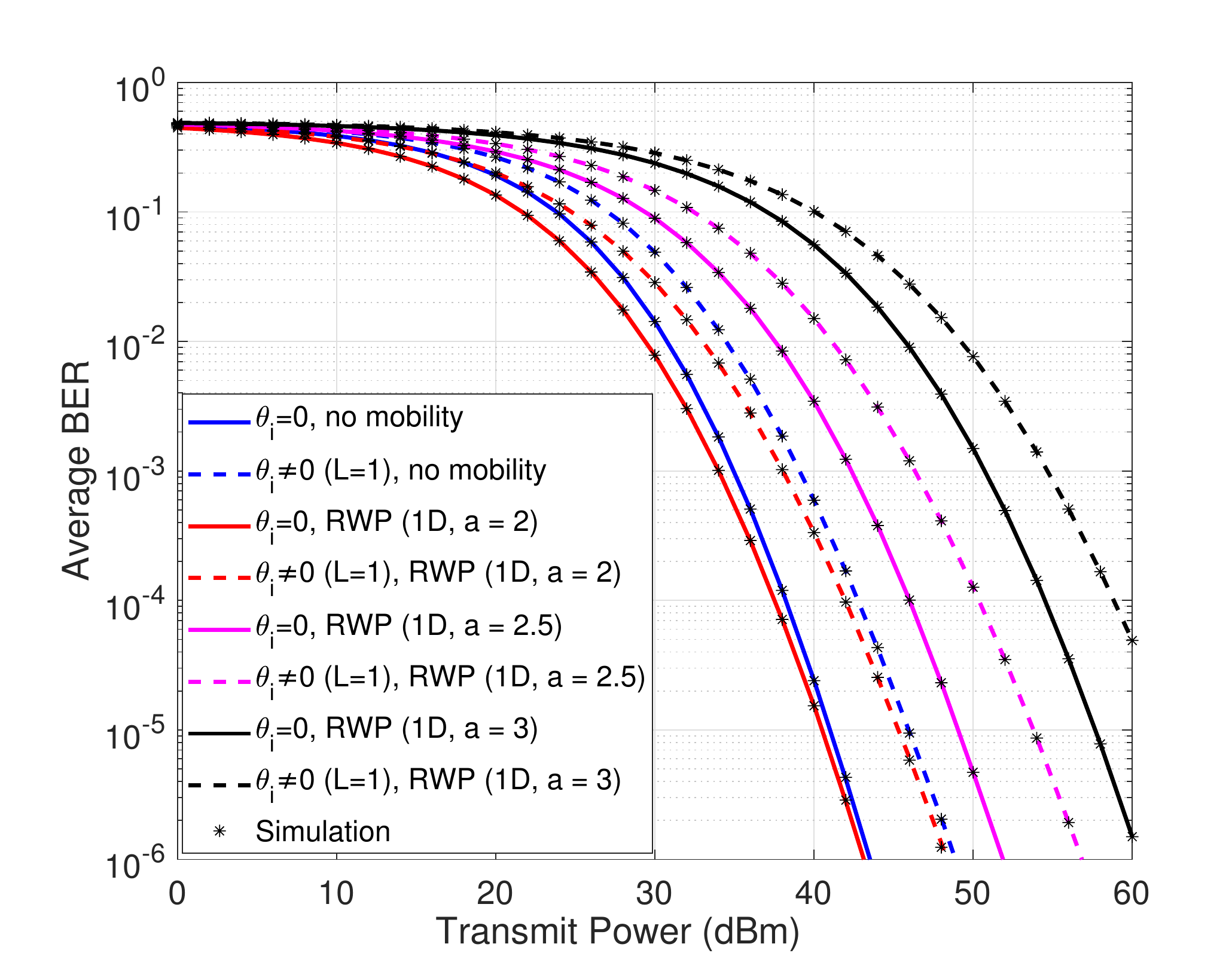}} 
	\caption{Average BER of the RIS-assisted system without direct link ($\omega=0$) for phase noise level $L=1$ at different mobility parameters.}
	\label{fig:ber_ris_mob}
\end{figure}

\begin{figure}
	{\includegraphics[scale=0.45]{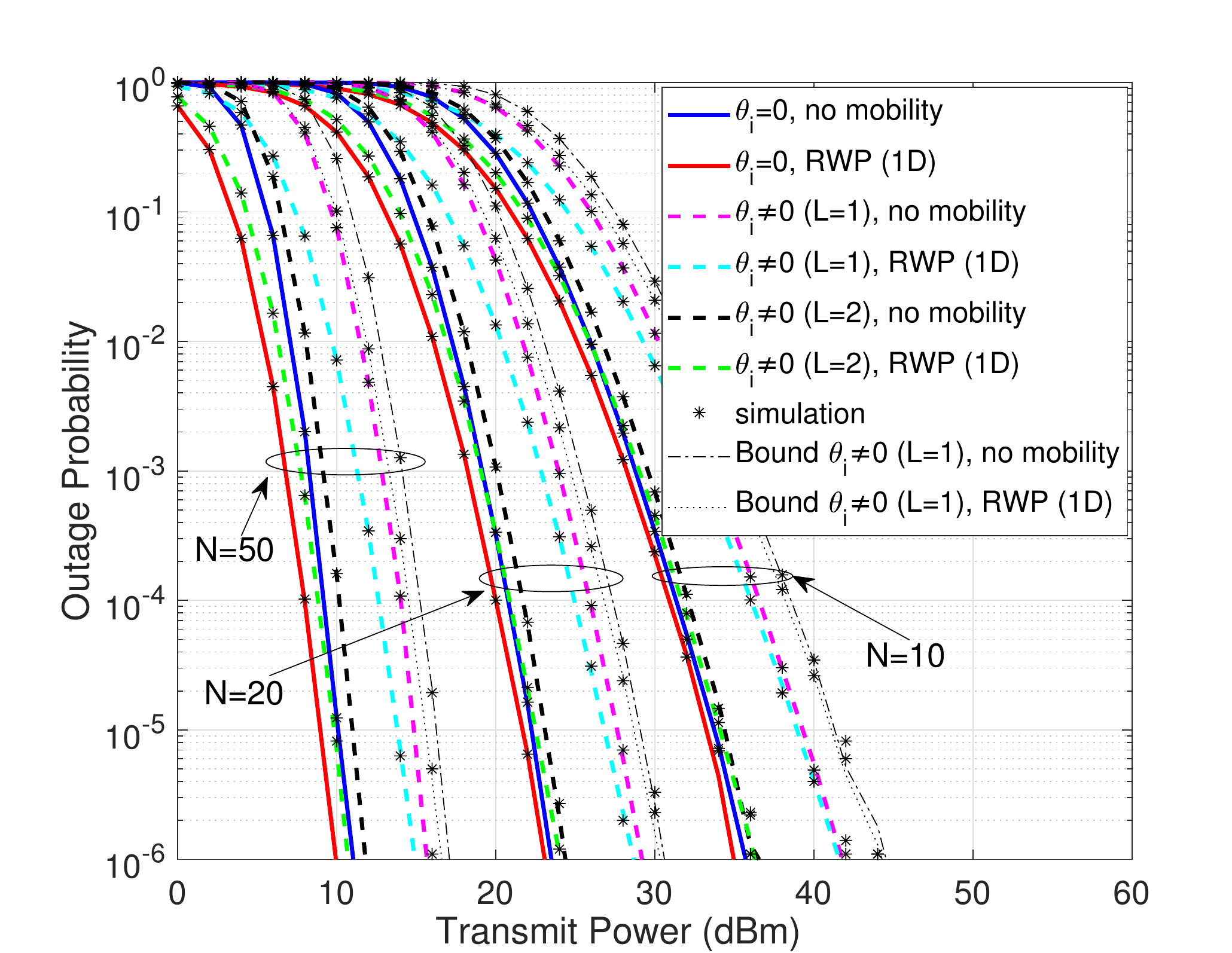}} 
	\caption{Outage performance of RIS-assisted system without direct link ($\omega=0$) with $1$-D mobility at different  phase noise parameters.}
	\label{fig:outage_ris_ph}
\end{figure}

\begin{figure}
	\centering
	{\includegraphics[scale=0.45]{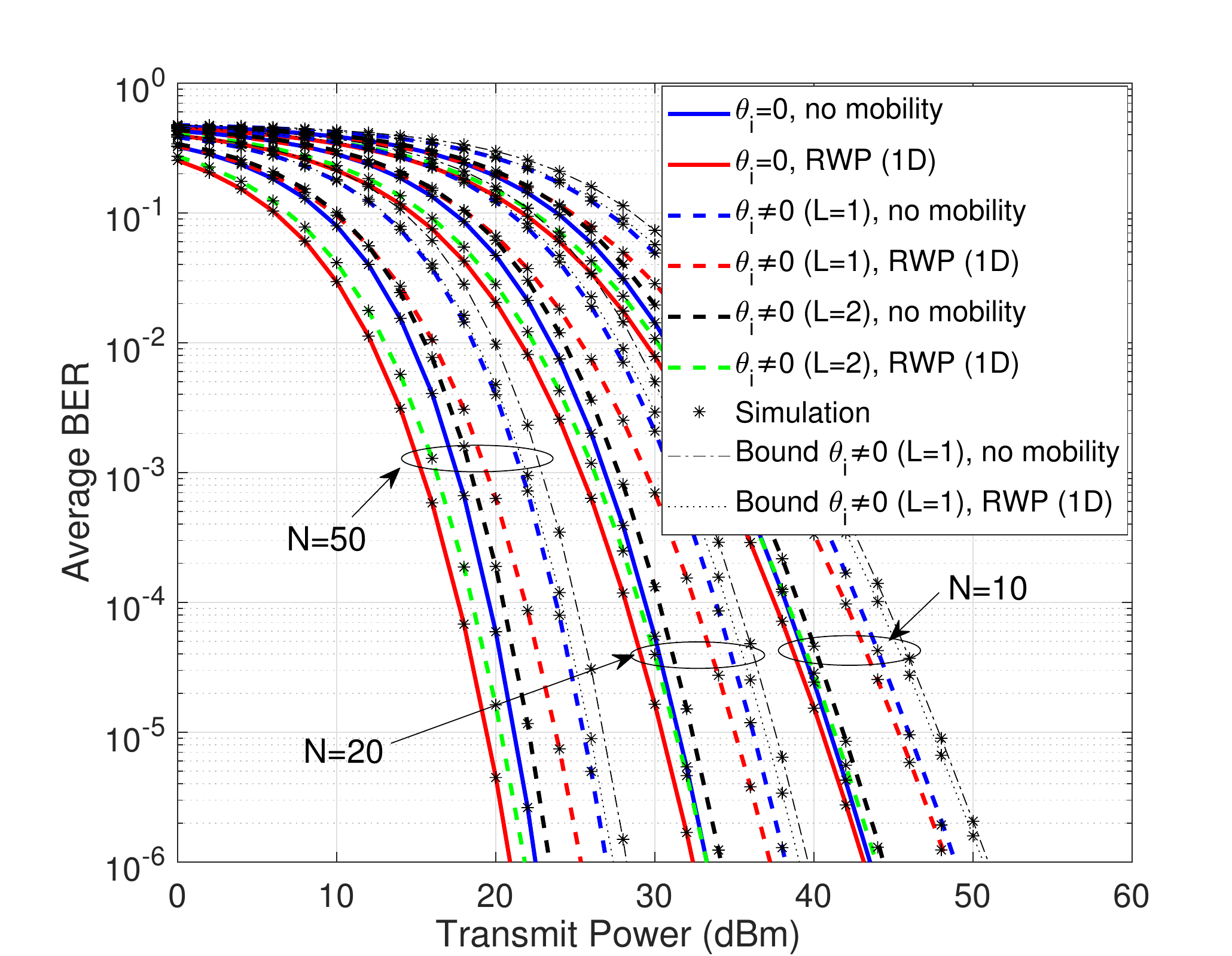}} 
	\caption{Average BER of RIS-assisted system  without direct link ($\omega=0$) with $1$-D mobility  at different  phase noise parameters.}
	\label{fig:ber_ris_ph}
\end{figure} 

\begin{figure}
	\centering
	{\includegraphics[scale=0.45]{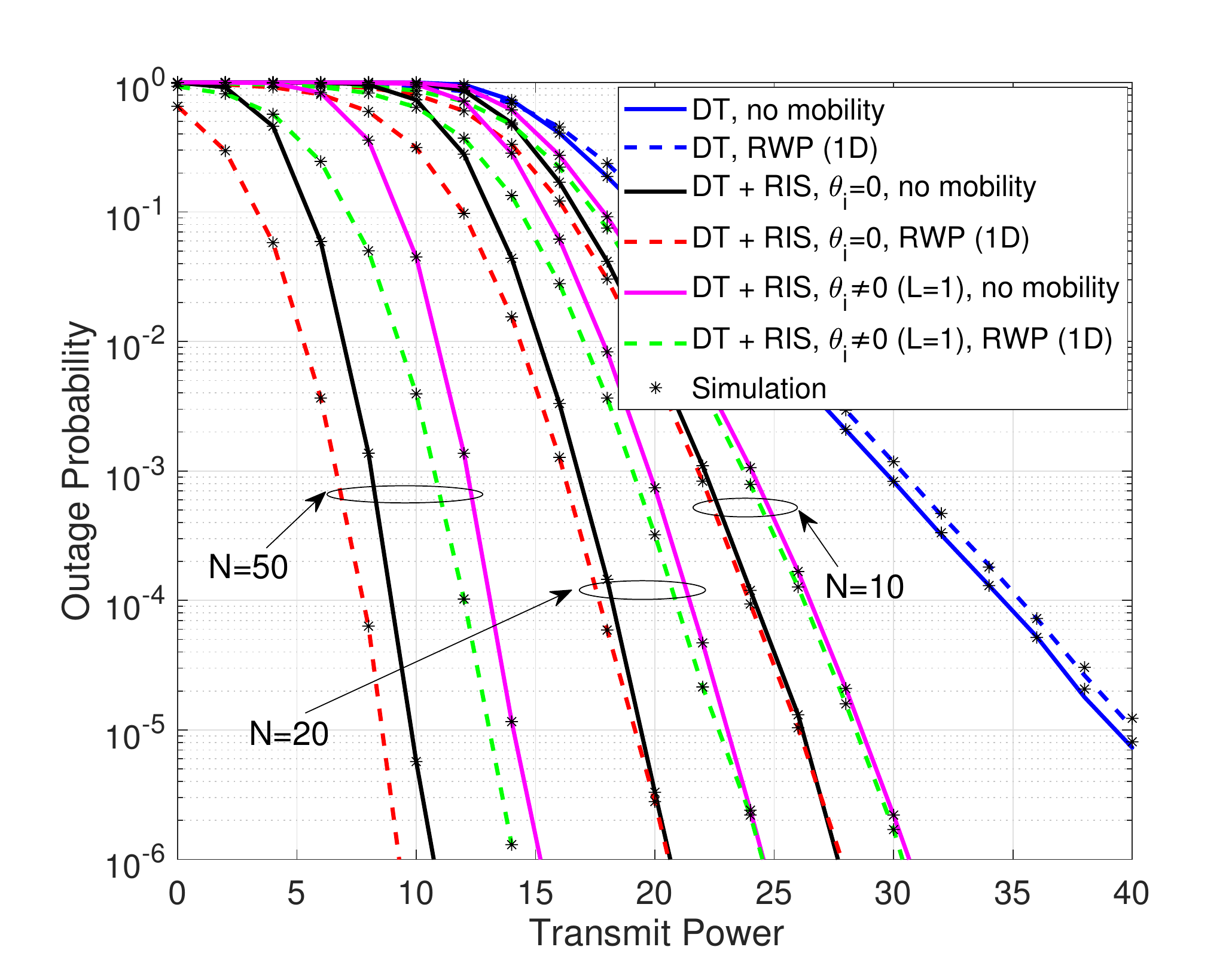}} 
	\caption{Outage Performance of  RIS-assisted system  with direct link ($\omega=1$) for different phase noise levels and varying mobility parameters.}
	\label{fig:outage_hd_ris}
\end{figure}

\begin{figure}
	\centering
	{\includegraphics[scale=0.45]{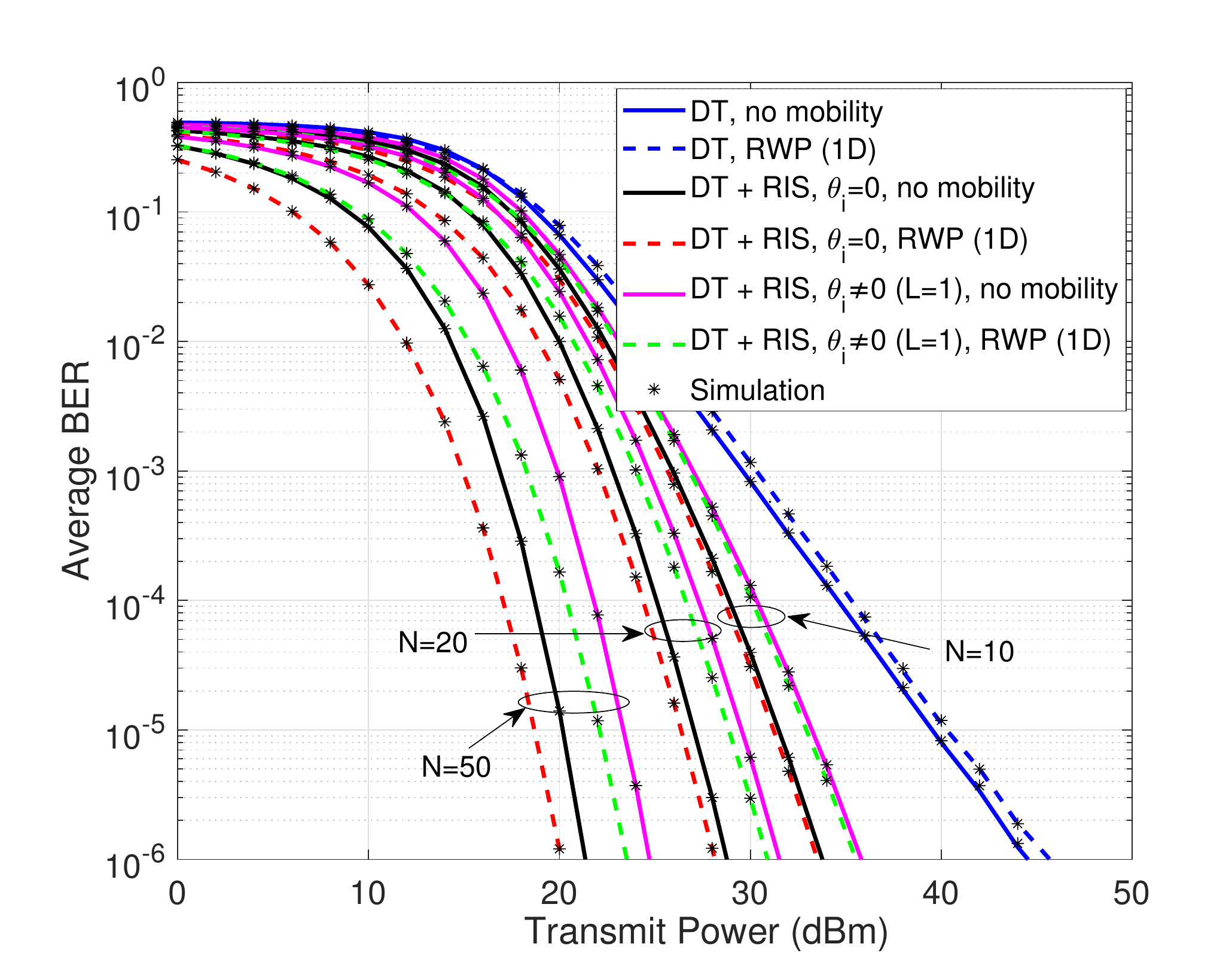}} 
	\caption{Effect of phase noise and varying mobility parameters on average BER of RIS-assisted system  with direct link ($\omega=1$).}
	\label{fig:ber_hd_ris}
\end{figure}

\section{Simulation and Numerical Results}\label{sec:sim_num_res}
In this section, we use numerical analysis and Monte Carlo simulations (averaged over $10^{8}$ channel realizations) to demonstrate the effect of phase noise and mobility on the performance of RIS-assisted transmission over generalized fading channels. We also  validate the derived analytical expressions and bounds using simulations.  For numerical computation, we use Python code implementation of multivariable Fox's H-function \cite{Alhennawi2016}. To demonstrate the effect of phase noise and mobility, we also plot the performance of RIS-assisted system with perfect phase compensation and no mobility for reference.  To compute the path loss component of the RIS without mobility, we assume that the user is placed at the half of the distance  $\frac{d_{2}}{2}$, and for the moving destination, we use the RWP model, as described in the system model section.

 We consider carrier frequency $f=6$\mbox{GHz}, transmit antenna gain $G_{T}=10$\mbox{dBi}, and receive antenna gain $G_{R}=10$\mbox{dBi}. We assume the RIS situated perpendicular to the source and destination  with a distance from the source to RIS at $d_1 = 50$\mbox{m} and the RIS to destination at $d_2=100$\mbox{m} such that direct transmission distance from the source to the destination  $d_d=\sqrt{d_{1}^{2}+d_{2}^{2}}=111.8$ \mbox{m}. A noise floor of $-74$\mbox{dBm} is considered over a $20$\mbox{MHz} channel bandwidth. For simulations, we assume i.i.d. channel fading coefficients from the source to all RIS elements and similarly from all RIS elements to the destination. Channel parameters used in the simulations are as follows: for the first link ($\kappa$-$\mu$ fading), we take  $\kappa_{i}=4$, $\mu_{i}=2$, and for the second link  (dGG fading), we take $\alpha_{i,1}=\alpha_{1}=2$, $\beta_{i,1}=\beta_{1}=1$, $\alpha_{i,2}=\alpha_{2}=2$ and $\beta_{i,1}=\beta_{2}=2$ $\forall i$. To model the phase noise, we consider two quantization levels $L=1$ and $l 2$, and assume different mobility models with varying path loss exponent factor $2\le a\le 5$ to demonstrate the effect of mobility on the systems performance.

\subsection{Effect of Phase Noise and Mobility without Direct Link $\omega=0$ (See Fig.~\ref{fig:outage_ris_mob}, Fig.~\ref{fig:ber_ris_mob},  Fig.~\ref{fig:outage_ris_ph},  and Fig.~\ref{fig:ber_ris_ph})}
 We demonstrate the outage probability and average BER considering a phase noise level $L=1$ (1-bit quantization) for different mobility scenarios and path loss exponent $a$, as shown  in  Fig.~\ref{fig:outage_ris_mob} and Fig.~\ref{fig:ber_ris_mob}, respectively. Fig.~\ref{fig:outage_ris_mob} shows the outage probability of RIS-assisted system for different mobility models with and without phase errors. We have also plotted the outage probability for the case when destination is not moving and is situated at the mid-way. Fig.~\ref{fig:outage_ris_mob} shows that  there is a difference in performance considering the case of with and without RWP model. The use of mobility model can provide a  more realistic performance estimate than the static case in the context of vehicular communications. Further, there is  difference of $1\mbox{dBm}$  and $2\mbox{dBm}$  in transmit power to achieve the same outage probability for $1$-D and $3$-D models, respectively.   Fig.~\ref{fig:outage_ris_mob}  also depicts the impact of imperfect phase compensation represented by $1$-bit quantization at RIS: a significant  $6\mbox{dBm}$  higher transmit power is required  to achieve a desired outage performance of $10^{-3}$ compared with the optimal phase compensating system. We also show the significance of adding  RIS elements $N$, which shows a significant scaling of the outage performance  for the RIS-assisted system.
 
 
 Fig.~\ref{fig:ber_ris_mob} shows the average BER of the RIS system for varying path loss exponent  for $N=10$, $L=1$ and $1$-D RWP model. Here, BER performance degrades with an increase in the path loss exponent $a$ from $2$ to $2.5$ and then to $3$, as expected. We can observe  about $8\mbox{dBm}$ loss in system performance when $a$ is changed from $2$ to $2.5$ and then $2.5$ to $3$. We have also plotted RIS performance  with phase errors but without effect mobility demonstrating the gap in the performance with an increase in the parameter $a$.
 
 In Fig.~\ref{fig:outage_ris_ph} and Fig.~\ref{fig:ber_ris_ph}, we demonstrate the impact of imperfect phase on the RIS-assisted system performance considering a fixed RWP mobility model ( $1$-D) with a path loss exponent $a=2$. We plot the  outage probability and average BER for different phase noise quantization levels $L=1$,  $L=2$ for various number of reflecting elements $N$ of the  RIS module, as shown in Fig.~\ref{fig:outage_ris_ph} and Fig.~\ref{fig:ber_ris_ph}, respectively. We compare the performance with an  optimal criteria assuming perfect phase compensation. The figure shows that the  performance for moving user is slightly better than the static user by comparing the optimal performance  with $1$-D mobility plots since the user can be near to the RIS using the RWP model compared to static user assumed  to be situated at the mid-point. Further, Fig.~\ref{fig:outage_ris_ph} shows  performance degradation in the presence of phase noise. It can be seen that a loss of $5\mbox{dBm}$  transmit power with $1$-bit quantization compared to $L=2$ to achieve an outage performance of $10^{-3}$. However, the effect of phase noise can be compensated by increasing  the quantization level $L$  or by increasing the number of RIS elements $N$.  Thus, the outage probability and average BER with a higher phase noise  quantization level $L=2$ or doubling $N$ at the fixed level $L=1$ achieves a near-optimal performance. It appears that higher order quantization steps can mitigate the effect of phase noise with higher complexity at the RIS, which is not desirable for RIS deployment. However, the figures show that an increase in  RIS elements $N$ can provide optimal  even at a lower quantization lewvel $L=1$ with large RIS size. For example, an average BER performance of $10^{-4}$ can be achieved with $N=10$, $L=2$ at $38\mbox{dBm}$ of transmit power. The same performance can  be achieved with higher $N=20$ at a lower quantization $L=1$ at a transmit power of $34\mbox{dBm}$. Further, a higher $N=50$ with $L=1$ require $24\mbox{dBm}$ of transmit power to achieve  an average BER performance of $10^{-4}$. Thus, a large-size RIS can provide near-optimal in the presence of  phase noise.
 
 In all the above figures, we also validate the derived analytical results by comparing numerical computations obtained from exact analytical expressions with Monte-Carlo simulations. Further, we have also plotted upper bound in Fig.~\ref{fig:outage_ris_ph} and Fig.~\ref{fig:ber_ris_ph}. The figure shows that exact  analysis has an excellent agreement with simulations but there is a difference between the exact and the upper bound  due to the use of arithmetic  and geometric mean inequality. Further, slope of  simulation plots  depict that the  diversity order of system is   independent of mobility and phase errors, which can be useful for adapting the system and channel parameters for optimized performance.

\subsection{Effect of Phase Noise and Mobility with Direct Link $\omega=1$ (See Fig.~\ref{fig:outage_hd_ris} and Fig.~\ref{fig:ber_hd_ris}) }

In this section, we demonstrate the performance of RIS-assisted with direct link $\omega=1$. For the direct link, we use thee generalized-$\mathcal{K}$ shadowing fading model with $m=1$, $\sigma_{dB}=4\mbox{dB}$ and $M=2.5454$. The outage and average BER performance of the RIS-assisted system combined with direct link are shown in Fig.~\ref{fig:outage_hd_ris} and Fig.~\ref{fig:ber_hd_ris}, respectively. It is clear form the plots that the combined system with direct link performs better than individual systems (direct link only and RIS-assisted without direct link $\omega=0$) for a wider range of  SNR. It can be seen that the direct  transmission (DT) performs better at low transmit powers  compared with a smaller $N$ RIS. However, RIS-assisted transmission (without direct link)  outperforms DT for a reasonable  $N$. Fig.~\ref{fig:ber_hd_ris} also demonstrates that the average BER performance of the combined system is significantly better than the direct transmission at a high transmit power by harnessing the line-of-sight signal  from the RIS elements. Moreover, the performance of RIS combined with direct transmission is always better than the RIS system (without DT) even at low SNR. For instance, we can observe a gain in transmit power of about $8\mbox{dBm}$ for $N=10$ and $14\mbox{dBm}$ for $N=20$ to achieve a desired outage performance of $10^{-3}$ compared to the DT. Hence, it is clear that combined system performance is better than individual systems thereby exploiting the presence of DT at low SNRs and the LOS signal received through RIS at high SNRs. further, in the presence of direct link, phase errors represented by lower quantization $L$ can be undone with proportionately lesser $N$ compared to RIS-assisted system without direct link.

\section{Conclusions}
We developed exact analysis and upper bounds  on the performance of RIS-assisted vehicular communication system considering phase noise with mobility over asymmetric  fading channels by coherently combining received signals reflected by RIS elements and direct transmissions from the source terminal.  We adopted the generalized uniform distribution to model the phase noise, the RWP model for mobility,  generalized-$K$ shadowed fading distribution for the direct link. We presented outage and average BER performance of the considered system in terms of univariate and multivariate Fox's H functions. We also derived simplified expressions at high SNR in terms of gamma functions to compute the diversity order of the system.  We used computer simulations to demonstrate the effect of phase noise and mobility on the RIS-assisted performance compared with the optimal performance with perfect phase compensation. For RIS-assisted vehicular network, consideration of a statistical model for mobility provides a better estimate on the performance. Further, the effect of phase noise can be mitigated by increasing the number of elements in the RIS module and using higher quantization level for the phase error. Further, our analysis demonstrates the effectiveness of the coherent combining  of reflected signals from RIS  and the signal from direct link  to achieve reliable performance even with phase noise and user under movement.

It would be interesting to analyze the system performance considering a measurement based mobility model and  correlated fading channels in the presence of phase noise.

\section*{Appendix A: PDF of Single-Element RIS with User Mobility and Phase Errors $Z_i$}
To compute the PDF  of $Z_i$ for the product of four random variable as $Z_{i}=\lvert h_{i}^{(f)}\rvert g_{l,i}^{\rm rwp}\lvert g_{i}^{(f)}\rvert e^{\J \theta_{i}}$, we first derive the PDF of second link with mobility $g_{i}=g_{l,i}^{\rm rwp}\lvert g_{i}^{(f)}\rvert$, then combine with the first link to get PDF of $Z_{hg,i}^{}= h_i g_i$, and finally apply a novel approach to derive the PDF $Z_i=Z_{hg,i}^{}e^{\J \theta_{i}} $.  

Assuming a  path gain model $g_{l,i}^{\rm rwp}=r^{-\frac{a}{2}}$, the PDF of dGG short-term fading combined with mobility model can be computed as 
\begin{eqnarray}\label{eq:comb_pdf}
	f_{g_{i}}(x) = \int_{0}^{d_{2}} f_{g_{i}^{}}(x/r) f_{r}(r) dr
\end{eqnarray}
where $f_r(r)$ is the PDF for the RWP model (see \eqref{eq:mobility_power_pdf}).
Thus, the PDF of $f_{g_{i}^{}}(x/r)$ for a given $r$ can be expressed as
\begin{eqnarray}\label{eq:dgg_pdf_mod}
	&f_{g_{i}^{}}(x/r) = r^{\frac{a}{2}} f_{\lvert g_{i}^{(f)}\rvert}(x r^{\frac{a}{2}}) = \frac{1}{\Gamma(\beta_{i,1})\Gamma(\beta_{i,2})} x^{-1} \nonumber \\ &H_{0,2}^{2,0} \Bigg[\begin{array}{c} \phi_{g,i}^{\frac{1}{\alpha_{i,2}}} r^{\frac{a}{2}} x^{} \end{array} \big\vert \begin{array}{c} - \\ (\beta_{i,2},\frac{1}{\alpha_{i,2}}),(\beta_{i,1},\frac{1}{\alpha_{i,1}})\end{array}\Bigg]
\end{eqnarray}
Substituting \eqref{eq:dgg_pdf_mod} in \eqref{eq:comb_pdf} and expanding the definition of Fox's H-function and interchanging the order of integration, we get
\begin{eqnarray}\label{eq:comb_pdf_2}
	&f_{g_{i}}(x) = \sum_{j_{i}=1}^{n} B_{j_{i}} \frac{1}{d_{2}^{\beta_{j}+1}} \frac{1}{\Gamma(\beta_{i,1})\Gamma(\beta_{i,2})} x^{-1} \frac{1}{2\pi \J} \int_{\mathcal{L}}   \phi_{g,i}^{\frac{s}{\alpha_{i,2}}}  \nonumber \\ &x^{s}  \Gamma(\beta_{i,2}-\frac{s}{\alpha_{i,2}}) \Gamma(\beta_{i,1}-\frac{s}{\alpha_{i,1}}) \int_{0}^{d_{2}} r^{\frac{as}{2}+\beta_{j_{i}}} dr ds
\end{eqnarray}

Invoking the inner integral in  \eqref{eq:comb_pdf_2} 
\begin{eqnarray}\label{eq:inner_int_1}
	\int_{0}^{d_{2}} r^{\frac{as}{2}+\beta_{j_{i}}} dr = \frac{\Gamma(\frac{as}{2}+\beta_{j_{i}}+1)}{\Gamma(\frac{as}{2}+\beta_{j_{i}}+2)} d_{2}^{\frac{as}{2}+\beta_{j_{i}}+1} 
\end{eqnarray}
and applying the definition of Fox's H-function, we represent  \eqref{eq:comb_pdf_2} 
\begin{eqnarray}\label{eq:comb_dgg_mob}
	&f_{g_{i}}(x) = \sum_{j_{i}=1}^{n}  \psi_{i,2} x^{-1} \nonumber \\&H_{1,3}^{2,1} \Bigg[\begin{array}{c} \zeta_{i,2} x^{} \end{array} \big\vert \begin{array}{c} (-\beta_{j_{i}},\frac{a}{2}) \\ (\beta_{i,2},\frac{1}{\alpha_{i,2}}),(\beta_{i,1},\frac{1}{\alpha_{i,1}}),(-1-\beta_{j_{i}},\frac{a}{2})\end{array}\Bigg] 
\end{eqnarray}
where $\psi_{i,2}=\frac{B_{j_{i}}}{\Gamma(\beta_{i,1})\Gamma(\beta_{i,2})}$, $\zeta_{i,2}=\phi_{g,i}^{\frac{1}{\alpha_{i,2}}} d_{2}^{\frac{a}{2}}$.

Next,  PDF of product of two random variables $Z_{hg,i}=\lvert h_{i}^{(f)}\rvert g_{i}$ can be derived as
\begin{equation}\label{eq:prod_pdf_proof_new_1}
	f_{Z_{hg,i}}(x) = \int_{0}^{\infty} \frac{1}{u} f_{\lvert h_{i}^{(f)}\rvert}(u)f_{g_{i}}(\frac{x}{u}) du
\end{equation}
Substituting \eqref{eq:hi_pdf_simp} and \eqref{eq:comb_dgg_mob} in \eqref{eq:prod_pdf_proof_new_1},  expanding the definition of Fox's-H function, and interchanging  the order of integration, we get
\begin{eqnarray}\label{eq:prod_pdf_proof_new_3}
	&f_{Z_{hg,i}}(x) = \sum_{k_{i}=0}^{\infty} \sum_{j=1}^{n} \psi_{i,1} \psi_{i,2} x^{-1} \frac{1}{2\pi \J} \int_{\mathcal{L}_{i}} (\zeta_{i,2}^{-1} x^{-1})^{s_{i}} \nonumber \\ &\frac{\Gamma(1+\beta_{j_{i}}-\frac{a}{2}s_{i}) \Gamma(\beta_{i,1}+\frac{1}{\alpha_{i,1}}s_{i}) \Gamma(\beta_{i,2}+\frac{1}{\alpha_{i,2}}s_{i}) }{\Gamma(2+\beta_{j_{i}}-\frac{a}{2}s_{i})} \nonumber \\ &\int_{0}^{\infty} u^{s_{i}+\mu_{i}+k_{i}-1} e^{-\zeta_{i,1} u} du ds_{i}
\end{eqnarray}
Solving the inner integral in \eqref{eq:prod_pdf_proof_new_3}
\begin{eqnarray}
	&\int_{0}^{\infty} u^{s_{i}+\mu_{i}+k_{i}-1} e^{-\zeta_{i,1} u} du \nonumber \\ &= \frac{1}{(\zeta_{i,1})^{s_{i}+\mu_{i}+k_{i}}}\int_{0}^{\infty} u^{s_{i}+\mu_{i}+k_{i}-1} e^{- u} du \nonumber \\ &= \frac{1}{(\zeta_{i,1})^{s_{i}+\mu_{i}+k_{i}}} \Gamma(s_{i}+\mu_{i}+k_{i})
\end{eqnarray}
and applying the definition of Fox's-H function, we can express \eqref{eq:prod_pdf_proof_new_3} as
\begin{eqnarray}\label{eq:prod_pdf_proof_new}
	&f_{Z_{hg,i}}(x) = \sum_{k_{i}=0}^{\infty} \sum_{j_{i}=1}^{n} \psi_{i} x^{-1} \nonumber \\&\hspace{0mm}H_{1,4}^{3,1} \bigg[\begin{array}{c}\zeta_{i,1} \zeta_{i,2}^{} x^{} \end{array} \big\vert \begin{array}{c}(-\beta_{j_{i}},\frac{a}{2}) \\ V_{i} \end{array}\bigg]
\end{eqnarray}
where $\psi_{i}=\psi_{i,1} \psi_{i,2} \zeta_{i,1}^{-(\mu_{i}+k_{i})}$, $V_{i}=(\beta_{i,2},\frac{1}{\alpha_{i,2}}),(\beta_{i,1},\frac{1}{\alpha_{i,1}}),(\mu_{i}+k_{i},1),(-1-\beta_{j_{i}},\frac{a}{2})$.

Finally, we apply a novel approach to develop  the PDF of $Z_{i}^{}=Z_{hg,i}e^{\J \theta_{i}}$, where $\theta_{i}\sim\mathcal_{U}(-q\pi,q\pi)$. Using the conditional expectation of random variables, we can get
\begin{equation}\label{eq:new_pdf}
	f_{Z_{i}^{}}(x) = \Aver{f_{Z_{i}^{}}(x/\theta_{i})} = \int_{-q\pi}^{q\pi} f_{Z_{i}^{}}(x/\theta_{i}) f_{\theta_{i}}(\theta) d\theta
\end{equation}
Using \eqref{eq:prod_pdf_proof_new},  the density function of $Z_{i}^{}$  given $\theta_{i}$:
\begin{eqnarray}\label{eq:new_pdf_1}
	&f_{Z_{i}^{}}(x/\theta_{i}) = e^{-\J \theta_{i}} f_{Z_{hg,i}}(\frac{x}{e^{\J \theta_{i}}}) = \sum_{k_{i}=0}^{\infty} \sum_{j_{i}=1}^{n} \psi_{i} x^{-1} \nonumber\\ &H_{1,4}^{3,1} \bigg[\begin{array}{c}\zeta_{i,1} \zeta_{i,2}^{} x^{} e^{-\J \theta_{i}} \end{array} \big\vert \begin{array}{c}(-\beta_{j_{i}},\frac{a}{2}) \\ V_{i} \end{array}\bigg]
\end{eqnarray}
We substitute \eqref{eq:new_pdf_1} in \eqref{eq:new_pdf} to get
\begin{eqnarray}\label{eq:new_pdf_2}
	&f_{Z_{i}^{}}(x) = \sum_{k_{i}=0}^{\infty} \sum_{j_{i}=1}^{n} \psi_{i} \frac{1}{2\pi \J} \int_{\mathcal{L}_{i}} (\zeta_{i,2}^{} \zeta_{i,1} )^{s_{i}} x^{s_{i}-1} \nonumber \\ &\frac{\Gamma(1+\beta_{j_{i}}+\frac{a}{2}s_{i}) \Gamma(\beta_{i,1}-\frac{s_{i}}{\alpha_{i,1}}) \Gamma(\beta_{i,2}-\frac{s_{i}}{\alpha_{i,2}}) }{\Gamma(2+\beta_{j_{i}}+\frac{a}{2}s_{i})} \nonumber \\ &\Gamma(\mu_{i}+k_{i}-s_{i})  \big(\Aver{e^{-\J \theta_{i} s_{i}}}\big) ds_{i}
\end{eqnarray}
Using $\Aver{e^{-\J \theta_{i} s_{i}}} = \frac{\sin{q\pi s_{i}}}{q\pi s_{i}} = \frac{1}{\Gamma(1-q s_{i}) \Gamma(q s_{i})} \frac{\Gamma(q s_{i})}{\Gamma(1+q s_{i})} = \frac{1}{\Gamma(1-q s_{i})\Gamma(1+q s_{i})}$ in \eqref{eq:new_pdf_2}, and applying the definition of Fox's H-function, we get \eqref{eq:prod_pdf_proof_phase}, which completes the proof of the theorem.

\section*{Appendix B: PDF and CDF of $Z_{\rm RIS}$}
The PDF for $Z_{\rm RIS}=\sum_{i=1}^{N}Z_{i}^{}$ can be computed using the MGF as $f_{Z_{\rm RIS}^{}}(z) = \mathcal{L}^{-1} \prod_{i=1}^{N} M_{Z_i^{}}(s)$, where
\begin{eqnarray}\label{eq:mgf_zi}
	&M_{Z_i^{}}(s) =\int_{0}^{\infty}e^{-sx} f_{Z_{i}^{}}(x) dz \nonumber \\
	&= \sum_{k_{i}=0}^{\infty} \sum_{j_{i}=1}^{n} \psi_{i}  \int_{0}^{\infty}x^{-1}e^{-sx} \nonumber \\ &H_{2,5}^{3,1} \bigg[\begin{array}{c}\zeta_{i,1} \zeta_{i,2}^{} x^{} \end{array} \big\vert \begin{array}{c}(-\beta_{j_{i}},\frac{a}{2}),(1,q) \\ V_{i},(0,q) \end{array}\bigg]  dx
\end{eqnarray}

We expand the Fox's H-function definition and interchange the order of integration \eqref{eq:mgf_zi} to get
\begin{eqnarray}\label{eq:mgf_zi_1}
	&M_{Z_i^{}}(s) = \sum_{k_{i}=0}^{\infty} \sum_{j_{i}=1}^{n} \psi_{i} \frac{1}{2\pi \J} \int_{\mathcal{L}_{i}} (\zeta_{i,2}^{} \zeta_{i,1} )^{s_{i}} \nonumber \\ &\frac{\Gamma(1+\beta_{j_{i}}+\frac{a}{2}s_{i}) \Gamma(\beta_{i,1}-\frac{s_{i}}{\alpha_{i,1}}) \Gamma(\beta_{i,2}-\frac{s_{i}}{\alpha_{i,2}}) }{\Gamma(2+\beta_{j_{i}}+\frac{a}{2}s_{i})} \Gamma(\mu_{i}+k_{i}-s_{i})  \nonumber \\ &\frac{1}{\Gamma(1-q s_{i})\Gamma(1+q s_{i})} \int_{0}^{\infty} e^{-sx} x^{s_{i,1}-1} dx ds_{i}
\end{eqnarray}

The inner integral in \eqref{eq:mgf_zi_1} can be solved using the identity \cite[3.381.4]{integrals} as $\int_{0}^{\infty} e^{-sx} x^{s_{i}-1} dx = s^{-s_{i}} \Gamma(s_{i})$, which can be used to develop MGF in terms of Fox's H-function. Since the  MGF of the sum is given as $M_{Z_{\rm RIS}}(s)=\prod_{i=1}^{N} M_{Z_i^{}}(s)$, the PDF of $Z_{\rm RIS}$ can be represented as
\begin{eqnarray}\label{eq:sum_pdf_1}
	&f_{Z_{\rm RIS}^{}}(x) = \mathcal{L}^{-1} \prod_{i=1}^{N} M_{Z_i^{}}(s) \nonumber \\ &= \sum_{k_{1},\cdots,k_{N}=0}^{\infty} \sum_{j_{1},\cdots,j_{N}=1}^{n} \prod_{i=1}^{N} \psi_{i} \frac{1}{2\pi \J} \int_{\mathcal{L}_{i}} (\zeta_{i,2}^{} \zeta_{i,1} s^{-1})^{s_{i}} \nonumber \\ &\frac{\Gamma(1+\beta_{j_{i}}+\frac{a}{2}s_{i}) \Gamma(\beta_{i,1}-\frac{s_{i}}{\alpha_{i,1}}) \Gamma(\beta_{i,2}-\frac{s_{i}}{\alpha_{i,2}})  \Gamma(\mu_{i}+k_{i}-s_{i})  \Gamma(s_{i})}{\Gamma(2+\beta_{j_{i}}+\frac{a}{2}s_{i})\Gamma(1-q s_{i})\Gamma(1+q s_{i})} \nonumber \\ &(\frac{1}{2\pi \J} \int_{\mathcal{L}} s^{-\sum_{i}^{N}s_{i}} e^{sx} ds) ds_{i}
\end{eqnarray}
Next, we apply the identity \cite[8.315.1]{integrals} to solve the inner integral as
\begin{eqnarray}\label{eq:sum_pdf_inner_int}
	\frac{1}{2\pi \J} \int_{\mathcal{L}} s^{-\sum_{i}^{N}s_{i}} e^{sx} ds = \Big(\frac{1}{x}\Big)^{-\sum_{i}^{N}s_{i}+1} \frac{1}{\Gamma(\sum_{i}^{N}s_{i})}
\end{eqnarray}
Finally, we use \eqref{eq:sum_pdf_inner_int} in \eqref{eq:sum_pdf_1} and apply the definition of N-multivariate Fox's H-function \cite[A.1]{M-Foxh} to get \eqref{eq:sum_pdf_pe_final}. Applying similar steps,  the  CDF  $Z_{\rm RIS}$ can be derived using $F_{Z_{\rm RIS}^{}}(x) = \mathcal{L}^{-1} \prod_{i=1}^{N} \frac{M_{Z_i^{}}(s)}{s}$, as given in \eqref{eq:sum_cdf_pe_final}.
\section*{Appendix C: PDF and CDF of Upper Bound $Y_{\rm RIS}$}
We use the identity \cite[2.8]{Kilbas_FoxH} to find the $r$-th moment of $Z_{i}^{}$:
\begin{eqnarray}\label{eq:prod_pdf_1}
	&\Aver{(Z_{i}^{})^{r}} = \int_{0}^{\infty} x^{r} f_{Z_{i}^{}}(x) dx \nonumber \\ &
	= \sum_{k_{i}=0}^{\infty} \sum_{j_{i}=1}^{n} \psi_{i} (\zeta_{i,2}^{} \zeta_{i,1} )^{-r} \Gamma(\mu_{i}+k_{i}+r) \nonumber \\ &\frac{\Gamma(1+\beta_{j_{i}}-\frac{a}{2}r) \Gamma(\beta_{i,1}+\frac{r}{\alpha_{i,1}}) \Gamma(\beta_{i,2}+\frac{r}{\alpha_{i,2}}) }{\Gamma(1+q r)\Gamma(1-q r)\Gamma(2+\beta_{j_{i}}-\frac{a}{2}r)} 
\end{eqnarray}
Substituting \eqref{eq:prod_pdf_1} in \eqref{eq:prod_pdf}, we get
\begin{eqnarray}\label{eq:prod_pdf_2}
	&f_{Y_{\rm RIS}^{}}(x) = \frac{x^{-1}}{2\pi \J} \int\limits_{\mathcal{L}} \prod_{i=1}^{N} \sum_{k_{i}=0}^{\infty} \sum_{j_{i}=1}^{n} \psi_{i} (\zeta_{i,2}^{} \zeta_{i,1} )^{-r} \nonumber \\ &\frac{\Gamma(1+\beta_{j_{i}}-\frac{a}{2}r) \Gamma(\beta_{i,1}+\frac{r}{\alpha_{i,1}}) \Gamma(\beta_{i,2}+\frac{r}{\alpha_{i,2}}) }{\Gamma(2+\beta_{j_{i}}-\frac{a}{2}r)} \nonumber \\ &\Gamma(\mu_{i}+k_{i}+r) \frac{1}{\Gamma(1+q r)\Gamma(1-q r)} x^{-r}dr
\end{eqnarray}
Using \eqref{eq:prod_pdf_2} in  \eqref{eq:pdf_approximation} and applying the definition of Fox's H-function, we can get an upper bound on the PDF $ f_{Z_{\rm RIS}^{}}(x)$  in \eqref{eq:pdf_ris_approx}.

Further, using the integral property, the CDF can be computed as $F_{Y_{\rm RIS}^{}}(x)=\int_{0}^{x} f_{Y_{\rm RIS}^{}}(u) du$:
\begin{eqnarray}\label{eq:prod_cdf_1}
	&F_{Y_{\rm RIS}^{}}(x) = \frac{1}{2\pi \J} \int\limits_{\mathcal{L}} \prod_{i=1}^{N} \sum_{k_{i}=0}^{\infty} \sum_{j_{i}=1}^{n} \psi_{i} (\zeta_{i,2}^{} \zeta_{i,1} )^{r} \nonumber \\ &\frac{\Gamma(1+\beta_{j_{i}}+\frac{a}{2}r) \Gamma(\beta_{i,1}-\frac{r}{\alpha_{i,1}}) \Gamma(\beta_{i,2}-\frac{r}{\alpha_{i,2}}) }{\Gamma(2+\beta_{j_{i}}+\frac{a}{2}r)} \nonumber \\ &\Gamma(\mu_{i}+k_{i}-r) \frac{1}{\Gamma(1-q r)\Gamma(1+q r)} \int_{0}^{x} u^{-1}u^{r} du dr 
\end{eqnarray}
Thus, using the  $I=\int_{0}^{x} u^{-1}u^{r} du=\frac{\Gamma(r)}{\Gamma(r+1)} x^{r}$ and substituting \eqref{eq:prod_cdf_1} in \eqref{eq:cdf_approximation}, we get \eqref{eq:cdf_ris_approx}.

\section*{Appendix D: PDF and CDF of  SNR with Direct Link}
Using the PDF of SNR in \eqref{eq_ref_zaf} and expanding the definition of multivariate Fox's H-function, we derive the MGF of $\gamma_{\rm RIS}$ as
\begin{eqnarray}\label{eq:mgf_ris_1}
	&M_{\gamma_{\rm RIS}}(s) = \sum_{k_{1},\cdots,k_{N}=0}^{\infty} \sum_{j_{1},\cdots,j_{N}=1}^{n} \prod_{i=1}^{N} \frac{\psi_{i}}{2} \nonumber \\ &\frac{1}{2\pi \J} \int_{\mathcal{L}_{i}} (\zeta_{i,2}^{} \zeta_{i,1} \sqrt{\frac{1}{\bar{\gamma}_{\rm RIS}}}^{})^{s_{i}} \nonumber \\ &\frac{\Gamma(1+\beta_{j_{i}}+\frac{a}{2}s_{i}) \Gamma(\beta_{i,1}-\frac{s_{i}}{\alpha_{i,1}}) \Gamma(\beta_{i,2}-\frac{s_{i}}{\alpha_{i,2}})  \Gamma(\mu_{i}+k_{i}-s_{i})  \Gamma(s_{i})}{\Gamma(2+\beta_{j_{i}}+\frac{a}{2}s_{i})\Gamma(1-q s_{i})\Gamma(1+q s_{i})} \nonumber \\ & \frac{1}{\Gamma(\sum_{i}^{N}s_{i})} \Big( \int_{0}^{\infty} \gamma^{\sum_{i=1}^{N}\frac{s_{i}}{2}-1} e^{-s\gamma} d\gamma\Big) ds_{i} 
\end{eqnarray}
where $\int_{0}^{\infty} \gamma^{\sum_{i=1}^{N}\frac{s_{i}}{2}-1} e^{-s\gamma} d\gamma =(\frac{1}{s})^{\sum_{i=1}^{N}\frac{s_{i}}{2}} \Gamma(\sum_{i=1}^{N}\frac{s_{i}}{2})$.

Similarly, to get the MGF for the SNR for direct transmission, we use \eqref{eq:comb_genk_mob} to get
\begin{eqnarray}\label{eq:mgf_dl_3}
	&M_{\gamma_{d}}(s) = \frac{1}{2\sqrt{\bar{\gamma}_{d} }} \sum_{j=1}^{n} B_{j} \frac{b^{M+m}d_{}^{\frac{a(M+m)}{2}}}{2^{M+m} \Gamma(m)\Gamma(M)} \nonumber \\ &(\sqrt{\frac{1}{\bar{\gamma}_{d}}})^{M+m-1} \frac{1}{2\pi \J} \int_{\mathcal_{L}_{d}} (\frac{b  d^{\frac{a}{2}}}{2} \sqrt{\frac{1}{\bar{\gamma}_{d}}}^{})^{s_{d}} \nonumber \\ &\frac{\Gamma(\frac{M-m}{2}-\frac{s_{d}}{2})\Gamma(\frac{m-M}{2}-\frac{s_{d}}{2}) \Gamma(1-\beta_{j}-\frac{a(M+m)}{2}+\frac{a}{2}s_{d})}{\Gamma(-\beta_{j}-\frac{a(M+m)}{2}+\frac{a}{2}s_{d})} \nonumber \\ &\Big( \int_{0}^{\infty} \gamma^{\frac{M+m}{2}+\frac{s_{d}}{2}-1} e^{-s\gamma} d\gamma\Big) ds_{d}
\end{eqnarray}
where the inner integral is given by $\int_{0}^{\infty} \gamma^{\frac{M+m}{2}+\frac{s_{d}}{2}-1} e^{-s\gamma} d\gamma =(\frac{1}{s})^{\frac{M+m}{2}+\frac{s_{d}}{2}} \Gamma(\frac{M+m}{2}+\frac{s_{d}}{2})$.
To get the PDF of $\gamma_{\rm RISD}$, we use $f_{\gamma_{\rm RISD}}(\gamma) = \mathcal{L}^{-1} \big(M_{\gamma_{\rm RIS}}(s) M_{\gamma_{d}}(s)\big)$ as
\begin{eqnarray}\label{eq:snr_pdf_4}
	&f_{\gamma_{\rm RISD}}(\gamma) = \sum_{k_{1},\cdots,k_{N}=0}^{\infty} \sum_{j_{1},\cdots,j_{N}=1}^{n} \sum_{j=1}^{n}  \prod_{i=1}^{N} B_{j} \frac{\psi_{i}}{4\gamma} \nonumber \\ &\frac{1}{\Gamma(m)\Gamma(M)} \frac{1}{2\pi \J} \int_{\mathcal{L}_{i}} (\zeta_{i,2}^{} \zeta_{i,1} \sqrt{\frac{\gamma}{\bar{\gamma}_{\rm RIS}}}^{})^{s_{i}} \nonumber \\ &\frac{\Gamma(1+\beta_{j_{i}}+\frac{a}{2}s_{i}) \Gamma(\beta_{i,1}-\frac{s_{i}}{\alpha_{i,1}}) \Gamma(\beta_{i,2}-\frac{s_{i}}{\alpha_{i,2}})  \Gamma(\mu_{i}+k_{i}-s_{i})  \Gamma(s_{i})}{\Gamma(2+\beta_{j_{i}}+\frac{a}{2}s_{i})\Gamma(1-q s_{i})\Gamma(1+q s_{i})} \nonumber \\
	& \frac{\Gamma(\sum_{i=1}^{N}\frac{s_{i}}{2})}{\Gamma(\sum_{i}^{N}s_{i})}    \frac{1}{2\pi \J} \int_{\mathcal_{L}_{d}} (\frac{b  d^{\frac{a}{2}}}{2} \sqrt{\frac{\gamma}{\bar{\gamma}_{d}}}^{})^{s_{d}} \nonumber \\ &\frac{\Gamma(M-\frac{s_{d}}{2})\Gamma(m-\frac{s_{d}}{2}) \Gamma(1-\beta_{j}-a(M+m)+\frac{a}{2}s_{d})}{\Gamma(-\beta_{j}-a(M+m)+\frac{a}{2}s_{d})} \Gamma(\frac{s_{d}}{2})  \nonumber \\ &(\frac{1}{\Gamma(\sum_{i=1}^{N}\frac{s_{i}}{2}+\frac{s_{d}}{2})}) ds_{i} ds_{d}
\end{eqnarray}
where we used the following expression:
\begin{equation}\label{eq:inner_eq_2}
	\int_{L} (s)^{-\sum_{i=1}^{N}\frac{s_{i}}{2}-\frac{s_{d}}{2}-\frac{M+m}{2}} e^{s\gamma}  ds = \frac{2 \pi \J \gamma^{\sum_{i=1}^{N}\frac{s_{i}}{2}+\frac{s_{d}}{2}+\frac{M+m}{2}-1} }{\Gamma(\sum_{i=1}^{N}\frac{s_{i}}{2}+\frac{s_{d}}{2}+\frac{M+m}{2})}
\end{equation}
Finally, we apply  the definition of $N$-multivariate Fox's H-function in \cite[A.1]{M-Foxh}, to get \eqref{eq:pdf_snr_final}.

To compute the CDF of SNR, we use eq. \eqref{eq:pdf_snr_final} in $F_{\gamma_{\rm RISD}}(\gamma) = \int_{0}^{\gamma} f_{\gamma_{\rm RISD}}(x) dx$ and expand $N$-multivariate Fox's H-function in terms of Mellin-Barnes integrals to solve the inner integral as 
\begin{eqnarray}\label{eq:cdf_inner_eq_2}
	I=\int_{0}^{\gamma} x^{\sum_{i=1}^{N}\frac{s_{i}}{2}+\frac{s_{d}}{2}-1}  dx=\frac{\Gamma(\sum_{i=1}^{N}\frac{s_{i}}{2}+\frac{s_{d}}{2})\gamma^{\sum_{i=1}^{N}\frac{s_{i}}{2}+\frac{s_{d}}{2}}}{\Gamma(\sum_{i=1}^{N}\frac{s_{i}}{2}+\frac{s_{d}}{2}+1)} 
\end{eqnarray} 
and we apply the definition of $N$-multivariate Fox's H-function to get \eqref{eq:cdf_snr_final}.

Similarly, to derive an upper bound for the PDF and CDF of resultant SNR, we use the upper bound for the  PDF of  $\gamma_{\gamma_{\rm RIS}}(\gamma)$ in \eqref{eq:pdf_ris_approx} such that
\begin{eqnarray}\label{eq:mgf_ris_approx_1}
	&M_{\gamma_{\rm RIS}}(s) \leq \frac{N}{2\gamma} \sum_{k_{1},\cdots,k_{N}=0}^{\infty} \sum_{j_{1},\cdots,j_{N}=1}^{n} \prod_{i=1}^{N} \psi_{i} \nonumber \\ & \frac{\Gamma(1+\beta_{j_{i}}-\frac{a}{2}s_{1}) \Gamma(\beta_{i,1}+\frac{s_{1}}{\alpha_{i,1}}) \Gamma(\beta_{i,2}+\frac{s_{1}}{\alpha_{i,2}}) }{\Gamma(2+\beta_{j_{i}}-\frac{a}{2}s_{1})} \frac{\Gamma(\mu_{i}+k_{i}+s_{1})}{\Gamma(1+q s_{1})\Gamma(1-q s_{1})}\nonumber \\ &   (\zeta_{i,1} \zeta_{i,2}^{} (\frac{1}{N}\sqrt{\frac{1}{\bar{\gamma}_{\rm RIS}}})^{N} )^{s_{1}}\Big( \int_{0}^{\infty} \gamma^{N\frac{s_{1}}{2}-1} e^{-s\gamma} d\gamma\Big) ds_{1} 
\end{eqnarray}
with the inner integral  as $\int_{0}^{\infty} \gamma^{\frac{Ns_{1}}{2}-1} e^{-s\gamma} d\gamma =(\frac{1}{s})^{\frac{Ns_{1}}{2}} \Gamma(\frac{Ns_{1}}{2})$.	


Using  \eqref{eq:mgf_ris_approx_1} and \eqref{eq:mgf_dl_3} in  $f_{\gamma_{\rm RISD}}(\gamma) \leq \mathcal{L}^{-1} \big(M_{\gamma_{\rm RIS}}(s) M_{\gamma_{d}}(s)\big)$ resulting into an inner integral: 
\begin{equation}\label{eq:snr_inner_int}
	\int_{L} (s)^{-N\frac{s_{1}}{2}-\frac{s_{d}}{2}} e^{s\gamma}  ds = \frac{2 \pi j \gamma^{N\frac{s_{1}}{2}+\frac{s_{d}}{2}-1} }{\Gamma(N\frac{s_{1}}{2}+\frac{s_{d}}{2})}
\end{equation}
Using the definition of bivariate Fox's H-function on the resultant expressing comprising \eqref{eq:mgf_ris_approx_1} and \eqref{eq:mgf_dl_3} with \eqref{eq:snr_inner_int}, we can get \eqref{eq:pdf_snr_final}. Similarly, the CDF can be computed as $F_{\gamma_{\rm RISD}}(\gamma) \leq \mathcal{L}^{-1} \big(\frac{1}{s}M_{\gamma_{\rm RIS}}(s) M_{\gamma_{d}}(s)\big)$ to get \eqref{eq:cdf_snr_final}, which concludes the proof of the theorem.	
\bibliographystyle{IEEEtran}
\bibliography{Multi_RISE_full}

\end{document}